\documentclass[12pt]{article}
\pdfoutput=1
\usepackage[latin1]{inputenc}

\usepackage{amsmath}
\usepackage{amsfonts}
\usepackage{amssymb}
\usepackage{graphicx}
\usepackage{geometry}
\usepackage{amssymb,epsfig,subfigure}
\usepackage{hyperref}
\usepackage{comment}

\usepackage[numbers,sort&compress]{natbib}

\def\simge{
    \mathrel{\rlap{\raise 0.511ex 
        \hbox{$>$}}{\lower 0.511ex \hbox{$\sim$}}}}

\def\simle{
    \mathrel{\rlap{\raise 0.511ex 
        \hbox{$<$}}{\lower 0.511ex \hbox{$\sim$}}}}

\makeatletter
\renewcommand\section{\@startsection {section}{1}{\z@}%
                                 {-3.5ex \@plus -1ex \@minus -.2ex}
                                   {2.3ex \@plus.2ex}%
                                   {\normalfont\large\bfseries}}
\renewcommand\subsection{\@startsection{subsection}{2}{\z@}%
                                   {-3.25ex\@plus -1ex \@minus -.2ex}%
                                     {1.5ex \@plus .2ex}%
                                     {\normalfont\bfseries}}
\renewcommand\subsubsection{\@startsection{subsubsection}{3}{\z@}%
                                   {-3.25ex\@plus -1ex \@minus -.2ex}%
                                     {1.5ex \@plus .2ex}%
                                     {\normalfont\itshape}}
\makeatother

\def\pplogo{\vbox{\kern-\headheight\kern -29pt
\halign{##&##\hfil\cr&{\ppnumber}\cr\rule{0pt}{2.5ex}&\ppdate\cr}}}
\makeatletter
\def\ps@firstpage{\ps@empty \def\@oddhead{\hss\pplogo}%
  \let\@evenhead\@oddhead 
}
\def\maketitle{\par
 \begingroup
 \def\thefootnote{\fnsymbol{footnote}}
 \def\@makefnmark{\hbox{$^{\@thefnmark}$\hss}}
 \if@twocolumn
 \twocolumn[\@maketitle]
 \else \newpage
 \global\@topnum\z@ \@maketitle \fi\thispagestyle{firstpage}\@thanks
 \endgroup
 \setcounter{footnote}{0}
 \let\maketitle\relax
 \let\@maketitle\relax
 \gdef\@thanks{}\gdef\@author{}\gdef\@title{}\let\thanks\relax}
\makeatother

\numberwithin{equation}{section}

\renewcommand{\dag}{\dagger}

\newcommand{\be}{\begin{equation}}
\newcommand{\bea}{\begin{eqnarray}}
\newcommand{\ee}{\end{equation}}
\newcommand{\eea}{\end{eqnarray}}
\newcommand\beq{\begin{equation}}
\newcommand\eeq{\end{equation}}

\newcommand{\mc}{\mathcal}

\renewcommand{\t}{\tilde}

\newcommand{\pp}{p_\parallel}

\newcommand{\qp}{q_\parallel}
\newcommand{\sv}{\text{sgn}(v)}

\def\be{\begin{equation}}
\def\ee{\end{equation}}
\def\ba#1\ea{\begin{align}#1\end{align}}
\def\bg#1\eg{\begin{gather}#1\end{gather}}
\def\bm#1\em{\begin{multline}#1\end{multline}}
\def\bmd#1\emd{\begin{multlined}#1\end{multlined}}

\def\({\left(}
\def\){\right)}
\def\[{\left[}
\def\]{\right]}

\begin{document}

\setcounter{page}0
\def\ppnumber{\vbox{\baselineskip14pt
}}
\def\ppdate{\footnotesize{SU/ITP-14/15}} \date{}

\author{Gonzalo Torroba and Huajia Wang\\
[7mm]
{\normalsize \it Stanford Institute for Theoretical Physics }\\
{\normalsize  \it Department of Physics, Stanford University }\\
{\normalsize \it Stanford, CA 94305, USA}\\
[3mm]}

\bigskip
\title{\bf  Quantum critical metals  in $4-\epsilon$ dimensions
\vskip 0.5cm}
\maketitle

\begin{abstract}
We study the quantum theory of a Fermi surface coupled to a gapless boson scalar in $D=4-\epsilon$ spacetime dimensions as a simple model for non-Fermi liquids (NFL) near a quantum phase transition. Our analysis takes into account the full backreaction from Landau damping of the boson, and obtains an RG flow that proceeds through three distinct stages. Above the scale of Landau damping the Fermi velocity flows to zero, while the coupling evolves according to its classical dimension. Once damping becomes important, its backreaction leads to a crossover regime where dynamic and static damping effects compete and the fermion self-energy does not respect scaling. Below this crossover and having tuned the boson to criticality, the theory flows to a $z=3$ scalar interacting with a NFL. 
We finally analyze the IR phases of the theory with arbitrary number of flavors $N_c$. When $N_c$ is small, the superconducting dome covers the NFL behavior; strikingly, for moderately large $N_c$ we find that NFL effects become important first, before the onset of superconductivity. A generic prediction of the theory is that the Fermi velocity and quasiparticle residue vanish with a power-law $\omega^\epsilon$ as the fixed point is approached. These features may be useful for understanding some of the phenomenology of high $T_c$ materials in a systematic $\epsilon$--expansion.
\end{abstract}
\bigskip

\newpage

\tableofcontents

\vskip 1cm

\section{Introduction}\label{sec:intro}

Understanding the dynamics of finite density quantum field theory (QFT) is a central problem in theoretical physics. A paradigmatic example is given by a Fermi surface interacting with gapless bosons, which underlies a wide range of systems in high energy and condensed matter physics. It can lead to a parametric enhancement of superconductivity and to the formation of new phases, and is believed to be relevant for the description of strongly correlated electron systems~\cite{Wenreview, Fradkinbook, Sachdevbook}. It also drives the dynamics of QCD at finite density, which exhibits rich phenomena in neutron stars and heavy ion collisions~\cite{Rajagopal:2000wf}. While there has been sustained progress over the last decades, a definite understanding of the possible phases and long distance dynamics of finite density QFT is still lacking.

In this work we study the coupled field theory of a massless scalar field and a finite density of fermions, with a Yukawa-type interaction $L\supset g \phi \psi^\dag \psi$. We do this in a controlled weak coupling expansion around the critical spacetime dimensionality $D=4-\epsilon$. Our goal is to perform a systematic analysis of the quantum effects in this theory, determine its renormalization group (RG) evolution, and describe the possible low energy phases. 

There are several phenomenological and theoretical reasons for undertaking this task. First, there is growing evidence that high $T_c$ superconductivity (SC) and non-Fermi liquid (NFL) behavior occur near quantum phase transitions, where interactions between bosonic excitations and finite density fermions become important. The theory of a gapless boson interacting with a density of nonrelativistic fermions is perhaps the simplest QFT example that can model this. It can also accommodate different generalizations, such as multiband contributions or anisotropies, some of which will be explored here.

The second motivation is to develop analytic approaches that can shed light on the strong dynamics present in many condensed matter systems of interest. Different nonperturbative techniques have been applied, including large $N$, field theory dualities and, more recently, holography. Here we will instead work near the critical 3+1 spacetime dimension where the theory is under perturbative control, and set up a systematic $\epsilon$--expansion (of which only the lowest order will be calculated). This has been successful in other areas of critical phenomena~\cite{Wilson:1971dc}, and we hope that it can also help understand strongly correlated electron systems. This idea, of course, has a long history, and related developments appear in~\cite{Alford:2001dt,norton-pincus, hertz, nortonNFL,NW1,NW2,Mross:2010rd,Dong:2012ua,Dalidovich:2013qta,Mahajan:2013jza,Fitzpatrick:2013rfa}. We will find that, already in the lowest order in $\epsilon$, some of the properties of the theory present striking similarities to phenomena that are believed to occur in strongly interacting systems. An encouraging example of this (see \S \ref{sec:phases}) will be a robust prediction of a NFL regime driving superconductivity and where the Fermi velocity and quasiparticle residue flow to zero as a power of the frequency $\sim \omega^\epsilon$, something which is observed in some high $T_c$ materials. 

The results of our analysis are very encouraging for describing the phenomenology of the cuprates, certain heavy fermion systems and iron-based superconductors. These materials appear to have quantum phase transitions where the Fermi liquid behavior breaks down, and a nontrivial interplay with superconductivity is observed. See~\cite{taillefer, shibauchi, coleman} for reviews and references to experimental results. Their phase diagram is believed to have a non-Fermi liquid coexisting with the superconducting dome, so it is very interesting that the theory considered in this paper can realize such a regime by varying the parameter $N_c$ in \S \ref{sec:phases}. Another direction where our approach may connect to experimental results is the recent measurement in YBCO that the effective mass $m^*$ is strongly enhanced as the quantum critical point is approached~\cite{effmass}. Here we find that in the perturbative fixed point in $d=3-\epsilon$ dimensions, the effective mass diverges as $m^* \sim 1/\omega^\epsilon$. The model considered in this paper then provides a controlled framework where phenomenological properties of strongly correlated materials may be understood.

Turning to more conceptual motivations, coupling a Fermi surface to a massless boson poses qualitatively new problems in the renormalization of non-relativistic QFT, which are absent in relativistic field theories or in Fermi liquids without the gapless boson. The main reason is that these modes have very different RG scalings that compete at the quantum level, making the analysis difficult and not fully understood. We will address this with the help of the $\epsilon$--expansion, which provides a formal RG prescription. However, doing so reveals (see \S \ref{subsec:vertex}) the existence of nonlocal counterterms in the theory, i.e. poles in $\epsilon$ whose residue is a singular function of frequency and momenta. We will offer only a partial resolution to this issue, which is currently under investigation~\cite{liamfuture}. The weakly coupled theory that we study in this work then serves to test RG approaches in a controlled setup, and can help to exhibit their limitations. It is an important open problem to develop a Wilsonian RG that extends~\cite{shankar, Polchinski:1992ed} to include a critical boson. Given our results below, this appears to be quite challenging (especially for gauge fields), and we hope that this work will motivate further developments in this direction.

Finally, a crucial ingredient that finite density brings in is the Landau damping of bosons due to the Fermi surface. This happens at a relatively high scale --one loop below $k_F$-- and in turn it leads to strong corrections on the dynamics of the quasiparticles. Previous approaches to this problem have focused on the long distance limit, using an overdamped form of the boson propagator with a $z=3$ dynamical exponent~\cite{hertz}. Our analysis will go beyond this limit, in that starting from the UV it incorporates the full form of Landau damping. This will allow us to study the approach to the Landau damping region from the perturbative theory, and will reveal a new crossover regime where static and dynamic damping effects compete. The complete RG flow will exhibit how the boson interpolates between the dynamical exponent $z=1$ in the UV and $z=3$ in the IR, and its corresponding backreaction on the Fermi surface.

\subsection{Outline}

Let us now provide an outline of our results. 
The first step is to perform a perturbative one loop RG analysis, which we present in \S \ref{sec:oneloop}. This requires calculating the fermion self-energy, vertex renormalization, and boson vacuum polarization. The first, presented in \S \ref{subsec:selfenergy}, exhibits a nonzero anomalous dimension that signals NFL behavior, and a running Fermi velocity $v/c \to 0$, something that was also observed in~\cite{Fitzpatrick:2013rfa}. The vertex correction is computed in \S \ref{subsec:vertex}, and has the nonlocal counterterm discussed above. We will extract from here a local contribution (proportional to the derivative of the self-energy) and will use it to renormalize the coupling. In this case, the one loop corrections exactly cancel, and the beta function is proportional to the classical dimension of the coupling (of order $\epsilon$). The theory in $4-\epsilon$ dimensions then does not admit a one loop fixed point, although, as we discuss below, a finite $N$ generalization will exhibit critical behavior.

The effects from Landau damping are studied in \S \ref{subsec:vacuumpol}. This analysis also explains how to tune the boson to criticality, something that will have important consequences on the low energy dynamics. After that, in \S \ref{sec:LD} and \S \ref{sec:IR}, we will develop an RG description that takes into account Landau damping and its backreaction on the fermions, and that interpolates between the UV perturbative regime and the deep IR.
Let us summarize the results of this analysis:
\begin{itemize}
\item Above the scale of Landau damping, the dynamics correctly reproduces the one loop results.
\item The effects of Landau damping become important in a way that depends on $v/c$. For slow fermions this occurs at a scale $M_D \sim g k_F/\sqrt{v}$. What follows is a new kind of crossover regime where static and dynamic screening effects are comparable, and that extends up to a parametrically lower scale $v M_D$. The fermion dispersion relation deviates from the usual logarithmic running or from a scaling form. For models with $v \gtrsim c$, this intermediate range collapses into a more direct crossover at the scale $M_D v^{-1/2}$. 
\item Both the fast and slow fermions then transit into the low energy overdamped regime, where perturbation theory reorganizes in terms of the effective coupling $g/\sqrt{v}$, and the flow of the Fermi velocity continues to $v/c \to 0$, albeit with a different slope than in the UV. The result is a $z\approx3$ boson coupled to the Fermi surface, with the interaction flowing to strong coupling. This is eventually cut off by a BCS enhanced instability.
\end{itemize}

Finally, in \S\ref{sec:phases} we study a generalization of the theory to include $SU(N_c) \times SU(N_f)$ nonabelian symmetries, where $N_c$ generalizes the $SU(2)$ spin, while $N_f$ can arise from additional channels in the electronic system.
Focusing directly on the overdamped regime, we find that the one-loop beta function changes sign for $N_c>1$, leading to
a NFL fixed point where the Fermi velocity and quasiparticle residue $Z$ have a power-law decay towards zero. Furthermore, by varying $N_c$ the model interpolates between a NFL completely covered by the SC dome, to the case where NFL effects become important before the onset of superconductivity. The theory could also have other instabilities and/or competing orders, and we leave a more detailed discussion of the IR phases to future work. 

\begin{figure}[h!]
\begin{center}
\includegraphics[width=0.8\textwidth]{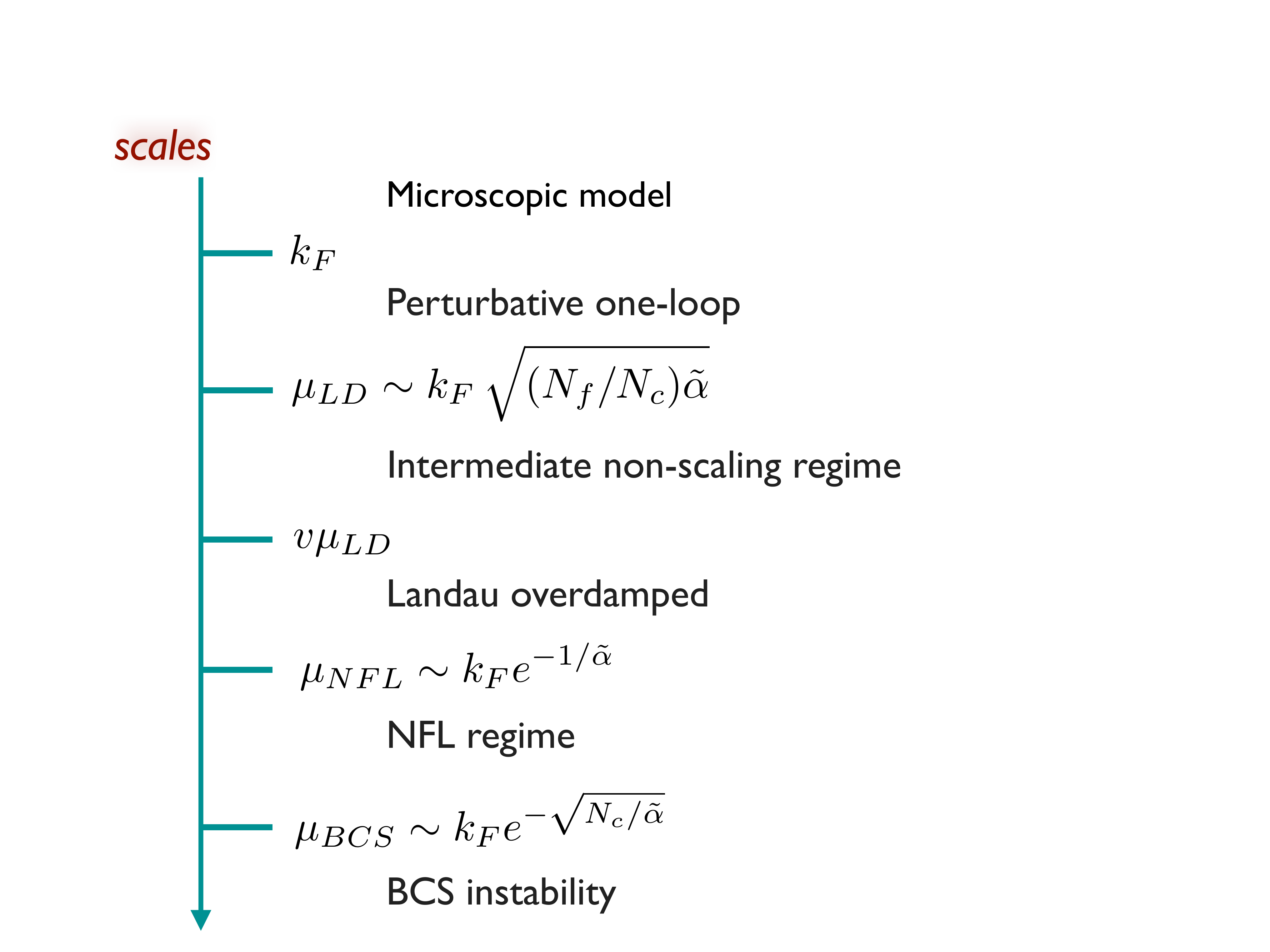}
\end{center}
\caption{\small{Energy scales for a Fermi surface coupled to a gapless boson. One loop perturbation theory is valid up to the Landau damping scale $\mu_{LD}$. For slow fermions $v/c \ll 1$ we find a large window of scales between $v \mu_{LD}$ and $\mu_{LD}$, where static and dynamic screening effects are equally important. This collapses to a rapid crossover for fast fermions. For small $N_c$ the approach to the NFL regime is stopped by the BCS instability. However, for moderately large $N_c$, the NFL regime sets in before the BCS instability, as in the Figure.}}\label{fig:scales1}
\end{figure}

The different energy regimes for the theory with $N_c$ `colors' and $N_f$ `flavors', with coupling $\tilde \alpha \equiv g^2 N_c/(12\pi^2v)$, are summarized in Figure \ref{fig:scales1}.

\section{Classical theory}\label{sec:classical}

Let us begin by defining the theory and reviewing its properties at tree level. We consider a critical boson interacting with a Fermi surface via a Yukawa type coupling in $D=d+1$ spacetime dimensions. We are interested in the case $d=3$, where the theory becomes weakly coupled, enabling a controlled perturbative expansion. We will also perform an $\epsilon$ expansion of the form $d=3-\epsilon$; this will be useful both as a regulator (a.k.a. dimensional regularization) and also to understand how the results here are modified as the physically important limit $d=2$ is approached.

The Euclidean action is
\be\label{eq:Sphi1}
S= \int d\tau\, d^dx \,\left\{\frac{1}{2} \left((\partial_\tau \phi)^2+c^2 (\vec \nabla \phi)^2 \right)+ \psi^\dag \left(\partial_\tau+ \varepsilon_F(i\vec \nabla)- \mu_F \right)\psi+ g_0 \phi \psi^\dag \psi \right\}\,.
\ee
Here $\mu_F$ is the chemical potential, $\varepsilon_F$ is the quasiparticle energy, and the Fermi surface is defined by $\varepsilon_F(\vec k_F)= \mu_F$, where $k_F$ is the Fermi momentum.  For our purpose of understanding the RG evolution and Landau damping backreaction, it will be enough to consider a spherical Fermi surface.
The details of $\varepsilon_F$ and the spin structure of $\psi$ depend on the specific model and UV completion of the theory. For instance, a massive Dirac fermion, at energies and chemical potential much smaller than the mass, gives an effective action for the particles of the form (\ref{eq:Sphi1}), with $\varepsilon_F(\vec k)= \frac{\vec k^2}{2m}$, while the antiparticles have energy $\omega \sim m$ and decouple. 

Let us discuss in more detail the interactions in this theory.
The Yukawa interaction written above can be generated, for instance, by decoupling a 4-Fermi interaction via a Hubbard-Stratonovich field. This in general gives a momentum-dependent coupling,
\be
S_{\psi-\phi}= \int d\tau\,d^dk\,d^dq\,g(k,q) \phi(q) \psi^\dag(k+q) \psi(k)\,;
\ee
the constant coupling that we use in the tree level action (\ref{eq:Sphi1}) arises from the limit of zero momentum transfer $q$ from the boson and zero Fermi surface angular momentum, $g_0 = g(|\vec k_F|, 0)$. We will find that one loop corrections generate a Yukawa coupling with strong momentum dependence. Furthermore, for simplicity in this work the $\phi^4$ interaction will be fine-tuned to vanish. While this is not important below the Landau damping scale ($\phi^4$ becomes irrelevant in that case), this coupling can lead to a richer RG flow in the UV limit. This has been recently studied in~\cite{sachdevfuture}.

\subsection{Spherical scaling}\label{subsec:spherical}

It is useful to study the system in terms of a spherical scaling towards the Fermi surface~\cite{shankar,Polchinski:1992ed}.\footnote{A different approach that is also often used in the literature is the ``patch scaling'' \cite{Polchinski:1993ii, NW1, NW2}. Also, a tree level RG analysis for the coupled boson-fermion system is given in~\cite{pepin}.} Here the fermion momentum is written in terms of a radial distance  $k_\perp$ towards the Fermi surface,
\be\label{eq:kdecomp}
\vec k = \hat n (k_F + k_\perp)\,,
\ee
where $\hat n$ is a $d$-dimensional unit vector, normal to the Fermi surface. In the low energy theory, $k_\perp \ll k_F$, and the fermion kinetic term is then of the form
\be
S_f= \int d\tau \frac{d \Omega_n}{(2\pi)^{d-1}} \frac{dk_\perp}{2\pi}\,\psi^\dag(\vec k) (\partial_\tau+ v k_\perp+ \frac{w}{2k_F} k_\perp^2 + \ldots) \psi(\vec k)\,.
\ee
Here $d\Omega_n$ is the volume element for the unit sphere parametrized by $\hat n$, and
$$
v= \varepsilon_F'(k_F)\;,\;w= k_F \varepsilon_F''(k_F)\,.
$$
Furthermore, the fermion was redefined to absorb an overall power $k_F^{d-1}$.
From now on, it will be convenient to set the boson speed $c=1$, and $v$ is in units of the boson velocity. For brevity, we refer to quadratic (and higher order) corrections to the fermion dispersion relation as ``curvature effects.''

Given the fermion momentum $\vec k$ in the direction $\hat n$, the boson momentum can be decomposed in components
\be
\vec p = \hat n p_\perp +\pp\,,
\ee
where $\pp$ is tangential to the Fermi surface. Its kinetic term becomes
\be
S_b= \int d\tau \frac{dp_\perp}{2\pi}\frac{d^{d-1}\pp}{(2\pi)^{d-1}}\,\frac{1}{2}\, \phi(-\vec p\,) (\partial_\tau^2 - p_\perp^2 - \pp^2) \phi(\vec p\,)\,,
\ee
In this spherical decomposition, the Yukawa interaction [recall (\ref{eq:kdecomp})]
\be
S_{int}=g \int d\tau\, \frac{d \Omega_n}{(2\pi)^{d-1}}\frac{dk_\perp}{2\pi} \frac{d^d p}{(2\pi)^d}\,\phi(\vec p\,)\,  \psi^\dag( \vec k+ \vec p\,) \psi(\vec k )\,,
\ee

Finally, let us consider the effect of a classical scaling transformation. As usual, the boson momentum scales towards the origin, $p_\mu'=e^b p_\mu$; however, the fermion scales towards the Fermi surface,
\be
k_0'=e^b k_0\;,\;k_\perp'=e^b k_\perp\,,
\ee
with the unit vector $\hat n$ fixed.  The action is classically invariant for
\be
\phi'(p')= e^{- \frac{d+3}{2}b} \phi(p)\;,\;\psi'(p')= e^{- \frac{3}{2}b} \psi(p)\;,\;g'= e^{\frac{d-3}{2}b} g\,.
\ee
The fermion scales as a two dimensional fermionic field (a consequence of the Fermi surface), while the boson retains its $d+1$ scaling dimension; the coupling becomes classically marginal at $d=3$, the dimension on which we focus.

\subsection{Renormalization and $\epsilon$--expansion}\label{subsec:epsilon}

Let us now study the $\epsilon$ expansion for $d=3-\epsilon$ space dimensions. The small parameter $\epsilon$ also provides a non-Wilsonian RG, very convenient especially for gauge theories~\cite{Bollini:1972ui,'tHooft:1972fi}. Setting $d=3-\epsilon$ implies here that the dimension of the Fermi surface is now formally $2-\epsilon$. 
Quantum corrections to the correlation functions will have poles as $\epsilon \to 0$, which are subtracted with counterterms in order to yield finite physical results. The dependence of the counterterms on $\epsilon$ can then be used to obtain the beta functions of the theory~\cite{Peskin:1995ev,ZinnJustin:2002ru}. Note that this is different from the $\epsilon$ expansion in~\cite{Dalidovich:2013qta}, where the analytic continuation is done on the codimension of the Fermi surface, instead of in the dimension. As a result, the low energy theory in both approaches will be different.

We begin from the original action (\ref{eq:Sphi1}) and denote its fields and couplings by a subindex `$0$'. These are the `bare' quantities, which are expressed in terms of counterterms and physical couplings as follows:
\be\label{eq:renormalized}
\psi_0= Z_\psi^{1/2} \psi\;,\;\phi_0= Z_\phi^{1/2} \phi\;,\;g_0= \mu^{\epsilon/2}\,\frac{Z_g}{Z_\phi^{1/2} Z_\psi}g\;,\;v_0= Z_v v\,,
\ee
where $g$ is dimensionless and $\mu$ is an arbitrary RG scale. The action becomes
\be\label{eq:Sphirenorm}
S= \int d\tau d^dx \left\{-\frac{1}{2} Z_\phi \phi \left(\partial_\tau^2+c^2 Z_c \vec \nabla^2 \right)\phi+Z_\psi  \psi^\dag \left(\partial_\tau+   Z_v \varepsilon(i\vec \nabla) \right)\psi+\mu^{\epsilon/2} Z_g g \phi \psi^\dag \psi \right\}\,.
\ee
At this stage we have kept the quasiparticle energy $\varepsilon(p)=\varepsilon_F(p)-\mu_F$ as a general spherically symmetric function that vanishes at $p =  k_F$. We will find below an interesting interplay between small $\epsilon$ and curvature effects from nonzero $\varepsilon''(p)$.

One consequence of the curvature of the Fermi surface is to renormalize the chemical potential, which is simply taken into account by a shift of the original $\mu_F$; we assume that this has been done in what follows, and don't write explicitly the required shift. Furthermore, the one loop boson self-energy is finite, so the scalar counterterms are not needed, $Z_\phi=Z_c=1$. It is also convenient to introduce
\be\label{eq:counter-def}
Z_\psi = 1+ \delta_\psi\;,\; Z_\psi Z_v= 1+ \delta_v\;,\;Z_g= 1+\delta_g\,.
\ee

The quantum corrections that have poles as $\epsilon \to 0$ are, as shown in \S \ref{sec:oneloop}, the fermion self-energy and vertex renormalization. Let us consider their effects. The inverse fermion propagator including counterterms and the self-energy $\Sigma$ is
\be
-G_F^{-1}(k_0, \vec k\,)= (i k_0 - \varepsilon(\vec k\,)) + \left( i \,\delta_\psi \,k_0- \delta_v \,\varepsilon(\vec k\,)\right)+\Sigma(k_0, \vec k\,)\,.
\ee
In minimal subtraction, $\delta_\psi$ and $\delta_v$ are chosen to cancel the poles of $\Sigma$. The vertex renormalization $\Gamma$ contributes to the cubic interaction as $L_{int}= \mu^{\epsilon/2} g(1+\delta_g +g^{-1}\Gamma) \,\phi\,\psi^\dag \psi$. The $\epsilon$ divergence of the vertex will turn out to have a nonlocal dependence on the boson momentum, and we discuss below in \S \ref{subsec:vertex} and \S\ref{subsec:RG1} a proposal for fixing $\delta_g$.

The last step is to obtain the beta functions. This is done by differentiating both sides of (\ref{eq:renormalized}) with respect to the arbitrary scale $\mu$, and noting that the bare couplings are independent of $\mu$. This gives the formulas for the fermion anomalous dimension, running velocity and coupling,
\bea\label{eq:scalarbetafc}
\gamma_\psi&=& \frac{1}{2}\, \mu \frac{d\delta_\psi}{d\mu} \nonumber\\
\beta_v&=&2 \gamma v -\mu\frac{d\delta_v}{d\mu} \\
\beta_g&=& \left(- \frac{\epsilon}{2} + 2 \gamma_\psi -\mu \frac{d\delta_g}{d\mu}\right)g\,.\nonumber
\eea
The counterterms in minimal subtraction depend on $\mu$ only through the running couplings; the derivatives in the above expression are then $\mu \frac{d \delta_X(g)}{d\mu}= \partial_g \delta_X\,\beta_g$, giving a system of equations that can be solved for the beta functions order by order in $1/\epsilon$. In what follows we will apply this renormalized perturbation theory to study one loop effects, and then to the theory including Landau damping effects.

\section{Perturbative analysis}\label{sec:oneloop}

\begin{figure}[h!]
\begin{center}
\includegraphics[width=0.80\textwidth]{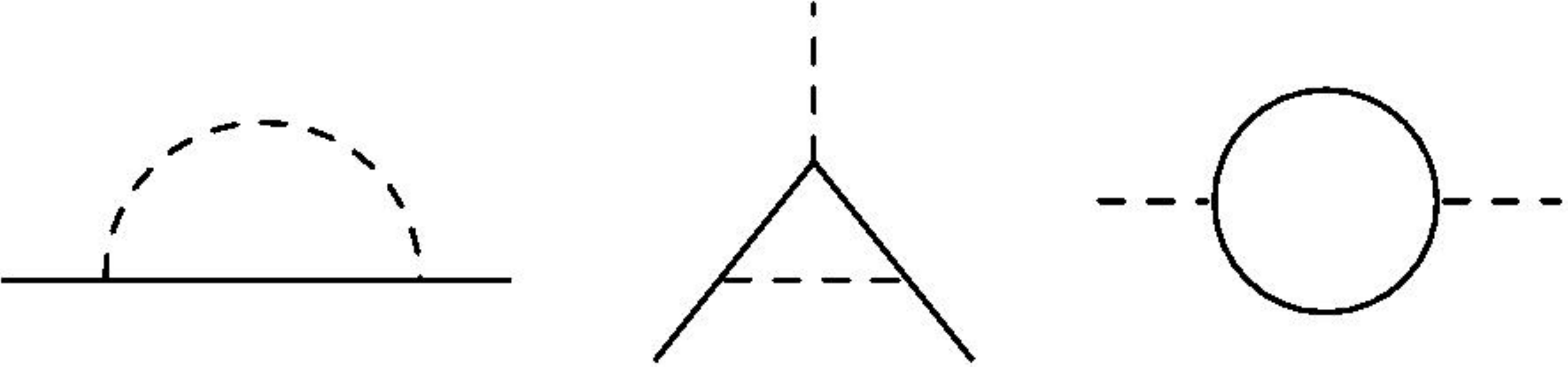}
\end{center}
\caption{\small{One loop corrections for the Fermi surface interacting with a scalar field. From left to right: fermion self-energy, vertex renormalization, and boson self-energy. The boson is represented by dashed lines.}}\label{fig:diagrams}
\end{figure}

We are now ready to study the quantum interactions between the Fermi surface and gapless boson, for which we will use the spherical scaling and $\epsilon$--expansion introduced in \S \ref{sec:classical}.
In this section, we present the one-loop calculations of these quantum corrections in the perturbative regime. These include the fermion self-energy $\Sigma$, the vertex correction $\Gamma$, as well as Landau damping (vacuum polarization) $\Pi$, which are shown in Figure \ref{fig:diagrams}. 
In the low energy theory, the first two have poles at small $\epsilon$ and determine the one-loop RG evolution; in contrast, the vacuum polarization is finite and does not affect the one loop beta functions. However, it dominates over the tree level frequency term over a large range of energies and momenta, and it has to be taken into account in order to obtain the correct IR physics. This will be done below in \S \ref{sec:LD}. There are also one loop effects that generate 4-boson and 4-fermion interactions, as can be seen in Figure \ref{fig:boxes}.
Details of the calculations can be found in the Appendix.

\begin{figure}[h!]
\begin{center}
\includegraphics[width=0.75\textwidth]{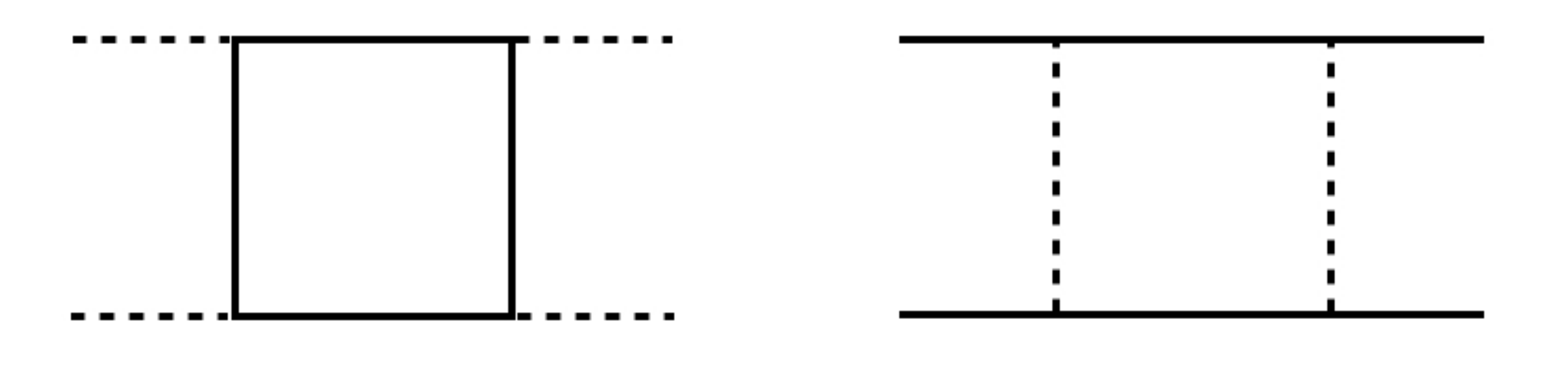}
\end{center}
\caption{\small{Some one loop corrections that induce $\phi^4$ and $\psi^4$ interactions. The boson is represented by dashed lines. The quartic boson correction is calculated in \S \ref{subsec:boxboson}, while $\psi^4$ in the BCS channel is related to the superconducting instability, discussed in later sections.}}\label{fig:boxes}
\end{figure}

Before presenting the results, it is important to clarify the origin of the UV divergences and running couplings that we will be calculating. At high energies and momenta (comparable to $k_F$), the one-loop corrections are made finite by the curvature of the Fermi surface; in particular, near 3+1 dimensions one finds that the self-energy and vertex depend logarithmically on $k_F$. These corrections will arise as UV divergences (or poles in $\epsilon$) here because we are focusing on the low energy theory very close to the Fermi surface. 

\subsection{Fermion self-energy}\label{subsec:selfenergy}

We begin by computing the one-loop fermion self-energy in the renormalized perturbation theory (\ref{eq:Sphirenorm}):
\be\label{eq:self-energy}
\Sigma(k_0, \vec k) =-g^2 \mu^\epsilon\,\int \frac{d^{D} p}{(2\pi)^{D}}\,\frac{1}{p_0^2+\vec p^{\,2}}\,\frac{1}{i(k_0+p_0)- \varepsilon(\vec k + \vec p)}\,,
\ee
with $\varepsilon(\vec p)= \varepsilon_F(\vec p)-\mu_F$.
In the limit when the external frequency and momentum $k_\perp$ are much smaller than the curvature of the Fermi surface, the fermion propagator can be linearized around the Fermi surface, obtaining
\be\label{eq:self-energy2}
\Sigma(k_0, \vec k) \approx-g^2 \mu^\epsilon\,\int \frac{dp_0\, d p_\perp \,d^{D-2} p_\parallel}{(2\pi)^{D}}\,\frac{1}{p_0^2+p_\perp^2+p_\parallel^2}\,\frac{1}{i(p_0+k_0)- v (p_\perp + k_\perp)}\,.
\ee
Here the momenta are defined by $\vec k = \hat n(k_F + k_\perp)$, $\vec p = \hat n p_\perp + \vec \pp$. The small corrections from the curvature of the Fermi surface will be discussed in \S \ref{subsec:curvature}.

In $D=4-\epsilon$ with small $\epsilon$, $\Sigma$ has an $\epsilon$ pole (plus finite terms), which determines the beta functions from (\ref{eq:scalarbetafc}). This has been computed before in~\cite{Fitzpatrick:2013rfa}, with the result (for external frequency and $k_\perp$ of the order of the RG scale $\mu$)
\be\label{eq:dimreg_sigma} 
\Sigma(k_0, \vec k)=\frac{g^2}{4\pi^2(1+|v|)}\left(i k_0 + \text{sgn}(v)k_\perp\right)\frac{1}{\epsilon} + \mc O(\epsilon^0)
\ee
and is also reproduced in the Appendix. The nonanalytic dependence on the Fermi velocity will play an important role below; $\sv$ is defined to vanish at $v=0$. In the theory including curvature effects, the discontinuous jump in $\sv$ is replaced by a smooth function of width controlled by $\mu/k_F$, where $\mu$ is the RG scale.

An important property of the self-energy is that it is not proportional to the tree level kinetic term $i k_0 - v k_\perp$, indicating both the usual wavefunction renormalization and that the Fermi velocity is also receiving quantum corrections.

\subsection{Vertex renormalization}\label{subsec:vertex}

Next, let us calculate the one loop vertex correction:
\be\label{eq:vertexoneloop1}
\Gamma(k;\,q)=\mu^\epsilon g^3 \int \frac{d^Dp}{(2\pi)^D}\,\frac{1}{(p_0-k_0)^2 + (\vec p- \vec k)^2}\,\frac{1}{i p_0 - \varepsilon(\vec p)}\,\frac{1}{i (p_0+q_0) - \varepsilon(\vec p+ \vec q)}\,,
\ee
where $k$ is the external fermion momentum and $q$ is the boson one. For external frequencies and momenta much smaller than $k_F$, we can again neglect quadratic and higher order terms in the fermion propagator, approximating
\be\label{eq:Vertex_correction}
\Gamma(k;q )\approx\mu^\epsilon\frac{g^3}{(2\pi)^D}\int\frac{dp_0\,dp_\perp\,d^{D-2}p_\parallel}{(k_0-p_0)^2+(k_\perp - p_\perp)^2+p_\parallel^2}\,\frac{1}{ip_0-v p_\perp}\,\frac{1}{i(p_0+q_0)-v(p_\perp+q_\perp)}\,.
\ee
Here the components of the momenta are given by $\vec k = \hat n(k_F+ k_\perp)$, $\vec p = \hat n(k_F+ p_\perp)+ \pp$, and $\vec q= \hat n q_\perp + \qp$.

Extracting the $\epsilon$ pole in $\Gamma$ is more nontrivial than for the self-energy. The reason is that there are now two contributions to the pole: one from the UV region, and another from a small low momentum region where the two fermionic poles approach the real axis. The calculation is performed explicitly in the Appendix, obtaining
\be\label{eq:dimreg_gamma}
\Gamma(k;q)=\frac{g^3}{4\pi^2}\frac{1}{1+|v|}\frac{iq_0+\text{sgn}(v) q_\perp}{iq_0-v q_\perp}\frac{1}{\epsilon} + \mc O(\epsilon^0)\,.
\ee

This result is quite striking: the Fermi surface interacting with a gapless boson has a 3-point function with an $\epsilon$ pole that depends nonlocally on $(q_0, q_\perp)$. The denominator in (\ref{eq:dimreg_gamma}), which is the same as that of the fermionic quasiparticles, suggests that this effect comes from integrating out light degrees of freedom near the Fermi surface. We note that nonlocal contributions in nonrelativistic QFTs have been observed before in e.g.~\cite{abanov,Brown:2000eh,Schafer:2004zf}.

In order to understand better the origin of this behavior, it is useful to relate the vertex to the fermion self-energy by a Ward-type identity (which is not exact in this theory). To derive it, the one loop expression
\be
\Gamma(k; q)=\mu^\epsilon\,g^3\int \frac{d^Dp}{(2\pi)^D}\, D(k-p)\,G_F(p)\,G_F(p+q)
\ee
(where $D$ and $G_F$ are the boson and fermion propagators) is multiplied on both sides by $iq_0 - v q_\perp$. 
For $q_0$ and $q_\perp$ much smaller than $k_F$, it is enough linearize the inverse propagators, $iq_0 - v q_\perp \approx G_F(p)^{-1}-G_F(p+ q)^{-1}$
and hence
\be\label{eq:ward_identity}
\left(iq_0 - v q_\perp\right)\Gamma(k; q)=g \left(\Sigma(k+q)-\Sigma(k) \right)\,.
\ee
Replacing here the one loop expression (\ref{eq:dimreg_sigma}) for $\Sigma$ reproduces (\ref{eq:dimreg_gamma}).

The diagrammatic Ward identity implies quite generally  that whenever the Fermi velocity runs (such that $\Sigma(q)$ is not proportional to $i q_0 - v q_\perp$) the vertex correction will have the singular dependence found in (\ref{eq:dimreg_gamma}). Although the identity (\ref{eq:ward_identity}) is not expected to be valid to all orders, this phenomenon is apparently more general. In the theory with a gauge field instead of a scalar, the same singular behavior is found, and in that case the Ward identity is a consequence of gauge invariance. This follows from the coupling of the gauge field to the Fermi surface, $L \supset \psi^\dag(\vec k+\vec q) V_\mu(\hat n) A_\mu(\vec q) \psi(\vec k)$, with $V_\mu=(i, - v\, \vec k/|\vec k|)$, and from the Ward identity of gauge invariance:
\be\label{eq:gauge_ward_identity}
q_\mu \Gamma_\mu(k;q) = g \left(\Sigma(k+q)-\Sigma(k) \right)\,.
\ee
Since at one loop $\Gamma_\mu \propto V_\mu$, we have
\be\label{eq:wardgauge}
\Gamma_\mu(k;q) = gV_\mu \,\frac{\Sigma(k+q)-\Sigma(k)}{V_\nu q_\nu}\,.
\ee
When the velocity runs, the numerator in this expression is no longer proportional to $V \cdot q$, leading to a momentum-dependent logarithmic divergence as in (\ref{eq:dimreg_gamma}).

The correct physical interpretation and consequences of (\ref{eq:dimreg_gamma}) are currently under investigation~\cite{liamfuture}, and here we only wish to make a few preliminary remarks. At first, (\ref{eq:dimreg_gamma}) is quite puzzling: renormalizing the theory would in principle require counterterms that depend on $q_0/q_\perp$, and similarly the RG flow would depend on this ratio. One point to note, however, is that the UV divergences of the low energy theory are actually IR effects in the microscopic model that describes the physics at scales above $k_F$. Indeed, in the theory (\ref{eq:Sphi1}), which includes the curvature of the Fermi surface, the poles in $\epsilon$ are replaced by $\log k_F$ factors. The interpretation of (\ref{eq:dimreg_gamma}) in this UV theory is then as a singular IR dependence of the 3-point function on external frequencies and momenta. Nevertheless, the appearance of these singular contributions casts doubt on the existence of a well-defined effective field theory description for the system, and more generally for nonrelativistic QFT.

Returning to the low energy theory near the Fermi surface, it is necessary to understand both the contribution of (\ref{eq:dimreg_gamma}) to the RG flow, and also its role in correlation functions. Let us consider the first point. The singular dependence on $(q_0, q_\perp)$ suggests that we have integrated out light degrees of freedom. This can be seen if we replace $\epsilon$ by a hard cutoff; analyzing the loop integral one finds contributions both from UV region and also from the IR ($p_0\sim q_0, |\vec{p}|\sim |\vec{q}|$), where the two fermion poles are on different sides of the real axis. The singular momentum dependence then comes from virtual low momenta particle-hole pairs.

Eq.~(\ref{eq:Vertex_correction}) reflects the non-Wilsonian nature of the $\epsilon$-expansion when applied to the Fermi surface interacting with a massless scalar, enhanced by a logarithmic divergence. In general we expect Wilsonian and non-Wilsonian approaches to have the same UV divergences; this is based on the intuition that the high momentum region of the loop integral dominates. When this holds, including the IR region in the integration does not change the leading UV dependence. However, here we see that this is not the case in the presence of a Fermi surface --the phase space suppression of the low momentum regime is compensated by an IR enhancement from the light quasiparticles, with the result that IR degrees of freedom also contribute to the UV divergence. 
It will be important to understand if a Wilsonian RG for the Fermi surface interacting with a gapless boson can be defined, a point to which we hope to return in the future. This seems challenging, especially given the relations (\ref{eq:ward_identity}) and (\ref{eq:wardgauge}) between a running Fermi velocity and the singular behavior of the vertex.

Independently of whether a consistent Wilsonian RG exists where the singular behavior of the vertex is resolved, (\ref{eq:Vertex_correction}) is the correct one-loop 1PI correction to the 3-point function. It will therefore be necessary to determine the effect of this singular correction on the physical observables and higher order correlation functions~\cite{liamfuture}. Somewhat similar issues arise in the proof of Migdal's theorem~\cite{AGD}, although in that case the vertex is not logarithmically enhanced. The expression (\ref{eq:Vertex_correction}) for the vertex also suggests a resonance between the quasiparticles and the boson if $v \sim c$. It will be interesting to understand the consequences of this and if, for instance, summation over soft modes can lead to similar divergences and also needs to be taken into account.

\subsection{Vacuum polarization and tuning to criticality}\label{subsec:vacuumpol}

The vacuum polarization for the boson (the last diagram in Fig.~\ref{fig:diagrams}) gives the familiar Landau damping of the scalar due to virtual particle-hole pairs. We will discuss the vacuum polarization in some detail, in order to determine how to tune the scalar to criticality.

The inverse boson propagator at one loop is (see the Appendix for more details)
\be
D^{-1}(p) = p_0^2+ \vec p^{\,2} + \Pi(p)+ \Pi_{ct}
\ee
where 
\be
\Pi(q)= - \mu^\epsilon g^2\int\,\frac{d^Dp}{(2\pi)^D}\,\frac{1}{i p_0 - \varepsilon(\vec p\,)}\,\frac{1}{i (p_0+q_0) - \varepsilon(\vec p+ \vec q\,)}\,,
\ee
and  $\Pi_{ct}$ is a constant counterterm that will adjust the boson to criticality (to be fixed below). For external frequencies and momenta smaller than $k_F$,
\be
\Pi(q)=-\mu^\epsilon\, \frac{g^2 k_F^{D-2}}{(2\pi)^D} \,\int \,\frac{dp_0 \,d p_\perp\, d^{D-2} \hat n}{\left(i p_0 - v p_\perp\right) \left(i(p_0+q_0)- v(p_\perp + \hat n \cdot \vec q\,) \right)}\,. \nonumber
\ee
This integral turns out to be convergent, but it depends on the order of integration. For $\epsilon \to 0$, it
evaluates to
\be\label{eq:Piscalar}
\Pi(q_0, \vec q\,) = - M_D^2 \left(C-\frac{q_0}{v | \vec q\,|}\,\tan^{-1} \frac{v| \vec q\,|}{q_0} \right)\,,
\ee
where the Debye scale
\be\label{eq:MD}
M_D^2 \equiv \frac{g^2 k_F^2}{2\pi^2 v}\,.
\ee
The constant $C$ depends on the ratio of the frequency and momentum cutoffs: $C=1$ for $\Lambda_{p_0} \gg \Lambda_{p_\perp}$, while $C=0$ in the opposite limit. This can also be checked using residues, which gives $C=1$ integrating over $p_0$ first, and $C=0$ integrating over $p_\perp$ first. We will discuss the consequences and interpretation of this UV ambiguity in a moment.

Before getting to this, let us discuss the unambiguous part of (\ref{eq:Piscalar}), namely the overall sign and the inverse tangent term. It will be useful to compare this result with the familiar Debye screening of the electrostatic potential $A_0$ in a charged Fermi liquid,
\be
\Pi_{00}(q)=M_D^2 \left(1-\frac{q_0}{v | \vec q\,|}\,\tan^{-1} \frac{v| \vec q\,|}{q_0} \right)\,.
\ee
$\Pi$ and $\Pi_{00}$ depend on the same dimensionless ratio $x\equiv\frac{q_0}{v|\vec{q}\,|}$ with the same functional form, but opposite in sign. Technically this comes from the extra factor of $i$ in the coupling $i A_0 \psi^\dag \psi$ as compared to $\phi \psi^\dag \psi$ in the Euclidean theory. Physically, this difference in sign reflects the repulsive and attractive nature of the force mediated by $A_0$ and $\phi$ respectively.

Now we need to understand how to fix $C$. As we just discussed, $C$ is ambiguous in the low energy theory, and it can be adjusted by using a constant counterterm $\Pi_{ct}$. To illustrate its effects, let us consider first $C=1$. In the static and dynamic limits
\be\label{eq:Piunstable}
\Pi(x \ll 1) \approx - M_D^2\;,\;\Pi(x \gg1) \approx - \frac{M_D^2}{3x^2}\,,
\ee
where $x= q_0/(v |\vec q\,|)$.
We learn that, while $\Pi$ is suppressed in the dynamic limit, it leads to an instability for nearly static fluctuations. In contrast, the static limit for the electrostatic potential would be, taking into account the sign difference with the real scalar, $\Pi_{00}(x \ll1) \approx M_D^2$; this positive mass squared is the familiar Debye screening of the Coulomb interaction. 

The instability means that the system should actually be in an ordered phase where the scalar condenses. In order to tune the boson to the critical point, from our low energy approach we will then choose $\Pi_{ct}$ to precisely cancel this contribution, leading to a critical boson with one loop vacuum polarization (see also~\cite{hertz})
\be\label{eq:Piscalarcrit}
\Pi(x)= M_D^2\,x\,\tan^{-1} \frac{1}{x}\;,\;x \equiv \frac{q_0}{v |\vec q\,|}\,.
\ee
The scalar is now Landau damped in the static limit, and screened in the dynamic regime:
\be\label{eq:Piscalarlimit}
\Pi(x \ll 1) \approx  \frac{\pi}{2} M_D^2\,|x|\;,\;\Pi(x \gg 1) \approx M_D^2\,.
\ee
For nearly on-shell bosons $|q_0|\sim |\vec q\,|$, slow fermions $v \ll c$ will then screen the scalar, while fast fermions $v \gg c$ produce a weaker Landau damping.
As an example, this is the expected behavior in systems where $\phi$ represents the magnetic order parameter.

More generally, we can parametrize the approach to the quantum critical point (QCP) by tuning a control parameter $u$ to its critical value $u_c$ giving, to leading order in $u-u_c$,
\be\label{eq:Pitunning}
\Pi(x)= M_D^2\,x\,\tan^{-1} \frac{1}{x}+ (u-u_c)+ \mc O((u-u_c)^2)\,.
\ee
In condensed matter systems $u$ can be e.g. doping or pressure. For $u<u_c$ the scalar tends to condense and the system orders, while for $u>u_c$ the boson becomes massive and we have a Femi liquid.

\subsection{Quartic boson vertex}\label{subsec:boxboson}

At one loop there is also a fermion ``box'' diagram that generates a $\phi^4$ interaction in the 1PI action, shown in Figure \ref{fig:boxes} above.
This diagram is finite and does not contribute to the one loop beta functions, but here we shall to briefly discuss it in order to illustrate the effects of the light quasiparticles on the scalar.

Let us compute the amplitude with external momenta ($p_\mu,q_\mu\rightarrow q_\mu, p_\mu$) for $\epsilon=0$:
\bea\label{eq:box_integral}
\Gamma^{(4)}(p,q) &=&- \frac{6g^4 k_F^2}{(2\pi)^4}\int dk_0 dk_\perp d^2\hat{n}\,\frac{1}{ik_0- v k_\perp}\,\frac{1}{i(k_0+p_0 )- v(k_\perp+\vec{p}\cdot\hat{n})}\nonumber\\
&&\frac{1}{i(k_0+p_0+q_0)- v(k_\perp+\vec{p}\cdot\hat{n}+\vec{q}\cdot\hat{n})}\,\frac{1}{i(k_0+q_0)- v(k_\perp+\vec{q}\cdot\hat{n})}\,,
\eea
where we have defined the vertex as a contribution $\Gamma^{(4)} \phi^4$ to the 1PI action.
This is analogous and related to the computation of vacuum polarization in the last section. Both vanish in a Wilsonian calculation, but the 1PI correlators are nonzero due to contributions from the region of low frequency and low momenta.
In particular, (\ref{eq:box_integral}) is also finite, yet ambiguous depending on the order of limits when doing the integration. In fact, by contracting two of the external legs, (\ref{eq:box_integral}) is identical to a two loop correction to the vacuum polarization.

We therefore resolve the ambiguity in the integration order in the same way as with the vacuum polarization. Based on the discussion in \S\ref{subsec:vacuumpol}, we shall follow the procedure of integrating over $k_\perp$ first followed by $k_0$, which produces the critical $C=0$ behavior of Landau damping in the last section.
The computation gives
\bea
\Gamma^{(4)}(p,q) &=&-i \frac{3g^4k_F^2}{4\pi^3 |v|}\int d^2\hat{n}\frac{1}{ip_0-v\vec{p}\cdot\hat{n}}\frac{1}{iq_0-v\vec{q}\cdot\hat{n}}\,\Bigg\{\frac{p_0+q_0}{i(p_0+q_0)-v(\vec{p}+\vec{q})\cdot\hat{n}}\nonumber\\
&&-\frac{p_0-q_0}{i(p_0-q_0)-v(\vec{p}-\vec{q})\cdot\hat{n}}\Bigg\}\,.
\eea
The effects of integrating out the fermions can be understood most simply if we take $\vec{p}\parallel \vec{q}$, for which the integral over the Fermi surface gives
\bea\label{eq:box_result}
\Gamma^{(4)}(p,q)  &=&\frac{3g^4k_F^2}{\pi |v|^3}\frac{1}{(x_p-x_q)^2}\Bigg(\frac{2}{q^2}x_p\tan^{-1}(x_p^{-1})+\frac{2}{p^2}x_q\tan^{-1}(x_q^{-1})\nonumber\\
&&-\left(\frac{1}{p}-\frac{1}{q}\right)^2x_{p-q}\tan^{-1}(x_{p-q}^{-1})-\left(\frac{1}{p}+\frac{1}{q}\right)^2x_{p+q}\tan^{-1}(x_{p+q}^{-1})\Bigg)
\eea
and we defined $x_p=\frac{p_0}{v|\vec p\,|}$, etc. 

As an example, consider the limit $q_0 \to 0$ first and then $\vec q \to 0$, relevant for the $z=3$ boson scaling ; the result is
\be
\lim_{\vec q \to 0} \,\lim_{q_0 \to 0}\,\Gamma^{(4)}(p,q)=-\frac{12g^4k_F^2}{\pi v^3}\,\frac{p_0^2}{(p_0^2+v^2 \vec p^{\,2})^2}\,.
\ee
This has a pole at the quasiparticle dispersion relation and, much as with the vertex, we find a resonance when the boson and fermion dispersion coincide. Here again, this is due to the non-Wilsonian contributions coming from the light Fermi surface excitations. While the $z=3$ exponent makes the boson self-interactions formally irrelevant below the Landau damping scale, it would be important to understand the effects of the momentum-dependent result (\ref{eq:box_result}), e.g. by resummation, since these cannot be ignored close to the Fermi surface. For planar systems this has been discussed in~\cite{abanov}.

\subsection{RG flow at one loop}\label{subsec:RG1}

We are finally in a position to determine the RG evolution of the theory at one loop.
Having computed the relevant quantum corrections, the counterterms are obtained by canceling the $\epsilon$ poles.  For the wavefunction and velocity counterterms we find (recall (\ref{eq:counter-def}))
\be
\delta_\psi= - \frac{g^2}{4\pi^2(1+|v|)}\,\frac{1}{\epsilon}\;,\;\delta_v= \frac{g^2\,\sv}{4\pi^2(1+|v|)}\,\frac{1}{\epsilon}\,.
\ee

Now we need to understand how to deal with the momentum-dependent $\epsilon$ pole of the vertex encountered in \S \ref{subsec:vertex}. We will take the somewhat conservative approach of only allowing momentum-independent beta functions; accordingly, we interpret the vertex (\ref{eq:dimreg_gamma}) as a sum of two contributions: a local term that renormalizes the Yukawa interaction, and a nonlocal renormalization for a different operator,\footnote{This was suggested by L. Fitzpatrick.}
\be\label{eq:Gamma-interpret}
\Gamma(k;q)= - i g \partial_{q_0} \Sigma(q) + \tilde \Gamma(k;q)\;,\;\tilde \Gamma(k;q)=\frac{g^3\,\sv}{4\pi^2 \epsilon}\,\frac{q_\perp}{iq_0- v q_\perp}\,.
\ee
Given this, the vertex counterterm
\be
\delta_g=- \frac{g^3}{4\pi^2(1+|v|)}\,\frac{1}{\epsilon}\,.
\ee
For a gauge field instead of a scalar, this is the familiar infinitesimal Ward identity, and in fact this is how usually the quantum vertex is defined. The full quantum corrections are however more complicated --there is an extra term $\tilde \Gamma$, whose implications are currently under investigation~\cite{liamfuture}.

Plugging these counterterms into (\ref{eq:scalarbetafc}) obtains the beta functions
\bea\label{eq:one-loop-beta}
\gamma_\psi = \frac{g^2}{8\pi^2(1+|v|)}\;,\;
\beta_v = \frac{g^2}{4\pi^2}\text{sgn}(v)\;,\;
\beta_g = -\frac{\epsilon}{2}g\nonumber\,.
\eea
Let us discuss the RG flow for $\epsilon=0$ first. The Yukawa coupling is marginal at one loop, and $\beta_v$ shows that there is an attractive fixed point where $v \to 0$, also described in~\cite{Fitzpatrick:2013rfa}. (Recall that $\sv$ vanishes at $v=0$). The velocity reaches its limiting value at a finite scale
\be\label{eq:vanishing_v}
\mu_{v=0}= e^{-4\pi^2 v_0/g_0^2} \Lambda\,,
\ee
where $g_0$ and $v_0$ are the values at $\mu=\Lambda$.
The non-Fermi liquid is characterized by an anomalous dimension $\gamma_\psi=g^2/(8 \pi^2)$. At finite but small $\epsilon$ the velocity still runs to zero, but this picture is corrected by a slow running of the Yukawa coupling. It would be interesting to compute the two loop correction to $\beta_g$ and see if they can lead to a perturbative fixed point.

\subsection{Curvature effects}\label{subsec:curvature}

So far we have studied the quantum theory in the low energy/momentum limit where the fermion dispersion relation becomes linear. Now we want to understand in more detail the effects from the nonzero curvature, and particularly its interplay with the $\epsilon$--expansion. For this, we include the leading correction to the linear term in the quasiparticle energy,
\be\label{eq:curvature1}
\varepsilon(\vec k\,)=\varepsilon_F(\vec k\,) - \mu_F= v k_\perp \left( 1+\frac{ c_1}{2} \frac{k_\perp}{k_F}+ \ldots \right)\,.
\ee
We will assume that $\varepsilon(\vec k\,)$ is a smooth function such that $c_1 = \frac{d \log \varepsilon'(k_F)}{d \log k_F}$ is some order one number. Recall that $v= \varepsilon_F'(k_F)$.

For concreteness, let us analyze the effect of the quadratic term in (\ref{eq:curvature1}) on the frequency-dependent part of the self-energy,
\be
-i g \partial_{k_0} \Sigma(k_0, k_\perp=0)=\mu^\epsilon g^2 \int \frac{d^Dp}{(2\pi)^D}\,\frac{1}{p_0^2 + \vec p^{\,2}}\,\frac{1}{\left[i(p_0+k_0)- \varepsilon(\vec p\,)\right]^2}\,,
\ee
which also contributes to the vertex renormalization. Including the quadratic term in $\varepsilon(\vec p\,)$ makes this integral convergent, in a way controlled by the scale $k_F$; we need to determine how the pole in $\epsilon$ arises in this context. 

It is convenient to integrate first over $p_0$ by residues. The external frequency is set to $k_0 \sim \mu$, the RG scale. Furthermore, $k_0$ can be ignored in the integrand if the lower integration range for $| \vec p\,|$ is taken from $k_0$; this is because $k_0$ is responsible for making the integral converge in the IR. With these simplifications, and changing to radial coordinates $p_\perp = p\,\cos \theta$, $\pp= p\,\sin \theta$, we have
\be
-i g \partial_{k_0} \Sigma\approx \frac{g^2}{8\pi^2} \frac{\mu^\epsilon}{k_F^\epsilon}\, \int_{\mu/k_F}^\infty \frac{dp}{p^{1+\epsilon}}\,\int_{-1}^1 d(\cos \theta) \,\frac{1}{\left(1+ v |\cos \theta+ c_1 p/2| \right)^2}\,.
\ee
The loop momentum has been redefined to absorb $k_F$, and the absolute value comes from the residue integral. Integrating over the angle and then over $p$ obtains, at small $\epsilon$,
\be\label{eq:Sigmacurvature}
-i g \partial_{k_0} \Sigma\approx \frac{g^2}{4\pi^2(1+|v|)}\,\frac{1}{\epsilon}\,\left(1- \frac{\mu^\epsilon}{k_F^\epsilon} \right)\,.
\ee
This result generalizes (\ref{eq:dimreg_sigma}) to include the curvature of the Fermi surface.

From this expression we can understand the interplay between the limits of small $\epsilon$ and large $k_F$. If we take the low energy limit $\mu/k_F \to 0$ at fixed (but small) $\epsilon$, then we recover (\ref{eq:dimreg_sigma}). This justifies our treatment so far of quantum effects ignoring the curvature of the Fermi surface. If, on the other hand, we take $\epsilon \to 0 $ first at fixed and small $\mu/k_F$, then (\ref{eq:Sigmacurvature}) gives
$$
-i g \partial_{k_0} \Sigma\approx \frac{g^2}{4\pi^2(1+|v|)}\,\log \frac{k_F}{\mu}\,,
$$
in agreement with the self-energy for the field theory with cutoff $\Lambda \sim k_F$ in 3+1 dimensions.

\section{Including backreaction from Landau damping}\label{sec:LD}

In the previous section we analyzed the theory of a Fermi surface interacting with a massless at the one loop level. The $D=4$ theory flows to an attractive fixed point $v/c \to 0$ with arbitrary coupling; this is corrected by a slow running of coupling for nonzero but small $\epsilon$. At one loop we also recovered the familiar Landau damping in the vacuum polarization $\Pi(x)$ of the scalar. Being a finite effect, it does not contribute to the one loop beta functions. However, the vacuum polarization becomes important at a scale controlled by $M_D$ in (\ref{eq:MD}), and at this point the one loop expansion (based on the tree level boson propagator) breaks down. This will in turn backreact on the fermionic sector, and can lead to a very different dynamics from that of \S \ref{sec:oneloop}. Our task is to set up a consistent RG treatment that incorporates these effects.

The traditional approach to this question has been to start from an IR effective theory where the boson propagator is approximated by 
\be\label{eq:IRD}
D^{-1}\approx \vec p^{\,2}+ \frac{\pi}{4} M_D^2 \frac{p_0}{v|\vec p|}\,.
\ee
This admits a $z=3$ dynamic critical exponent around which one could try to build a scaling theory~\cite{hertz}; see also~\cite{Sachdevbook} for a review and references to the original works. There are, however, two main concerns here. First, is this approach (which is different from a perturbative loop expansion) self-consistent? Furthermore, starting from the perturbative description in the UV, does the theory flow into this regime and, if so, how is the interpolation done?

In order to answer these questions, we will use the one loop resummed boson propagator including the full Landau damping to compute quantum effects on the Fermi surface. Summing the geometric series of one loop corrections gives the corrected propagator
\be\label{eq:Ddamped}
D^{-1}(p) = p_0^2+ \vec p^{\,2} + \Pi(p)\;,\;\Pi(p)= M_D^2\,\frac{p_0}{v|\vec p\,|}\,\tan^{-1} \frac{v |\vec p\,|}{p_0}
\ee
and it will be important to keep the complete vacuum polarization $\Pi(p)$. This goes beyond previous treatments in that we keep the full one loop resummed propagator, and not just an IR approximation, and will allow us to determine how the UV and IR limits are connected by RG. In the process we will uncover a novel crossover regime between these two limits, which for small $v$ can be made parametrically large and arises from competing static and dynamic screening effects.

\subsection{Basic aspects of Landau damping}\label{subsec:Piresum}

Let us discuss the relevant scales in the system, and
justify the resummation procedure in more detail. From the one loop results, and taking for simplicity $\epsilon=0$ so that the coupling doesn't run, the scale at which non-Fermi liquid effects become important is of order
\be\label{eq:NFL}
\mu_{NFL}= e^{-4\pi^2(1+v_0)/g_0^2}\Lambda\,,
\ee
where $v_0$ and $g_0$ are the values at the cutoff scale $\Lambda \sim k_F$. Around this scale, the fermion anomalous dimension contribution is comparable to the tree level kinetic term. Similarly, the scale at which the Fermi velocity $v \to 0$ is
\be\label{eq:vanishing_v1}
\mu_{v=0}= e^{-4\pi^2 v_0/g_0^2} \Lambda\,.
\ee

We want to compare these to the scale $\mu_{LD}$ when Landau damping $\Pi(p)$ becomes comparable to the tree level boson propagator. Since $\Pi(p)$ depends on the ratio $p_0/|\vec p\,|$, the strength of the vacuum polarization correction can vary in different kinematic regimes, as shown in (\ref{eq:Piscalarlimit}).
A simple way to define $\mu_{LD}$ is to require that for a nearly on-shell boson $\Pi$ becomes comparable to the tree level terms. This gives
\be\label{eq:muLD}
\mu_{LD} \approx \left(v^{-1}\,\tan^{-1}(v) \right)^{1/2}\, M_D\,.
\ee
Here the parameters on the right hand side are the physical couplings, which have to be evaluated at $\mu_{LD}$ according to the beta function runnings; so this is a self-consistent equation for the physical Landau damping scale. We will see momentarily that the running of the couplings will not play a crucial role in the regime of interest, so to a good approximation we can use their values at the UV cutoff $\Lambda$ to evaluate $\mu_{LD}$.

We will be interested in calculating the backreaction of Landau damping on the Fermi surface. For this we need to insert the corrected boson propagator into the fermion self-energy diagram, and find which regions of loop energy and momenta dominate the integral. We will find that the corrected self-energy deviates appreciably from (\ref{eq:dimreg_sigma}) at a scale which is also of order (\ref{eq:muLD}), so the estimate of Landau damping effects by setting the boson to its mass shell will turn out to be a good approximation. Another point to stress is that Landau damping effects are quite different for slow and fast fermions:
\be\label{eq:muslow-fast}
\mu_{LD}(v \ll 1) \approx M_D\;,\;\mu_{LD}(v \gg 1) \approx \frac{M_D}{v^{1/2}}\,,
\ee
reflecting the static and dynamic limits (\ref{eq:Piscalarlimit}) of the damping factor. Therefore, $v \gg 1$ tends to suppress Landau damping, while $v \ll 1$ has the opposite effect (recall that $M_D^2 \propto 1/v$).

The last important scale in the problem comes from the superconducting instability. This is enhanced by the presence of the gapless boson, and is of order~\cite{Son:1998uk},
\be\label{eq:gapscale}
\Delta \sim e^{-\gamma \pi^2 v_0^{1/2}/g_0} \Lambda
\ee
and $\gamma \sim \mathcal O(1)$. Therefore, at weak coupling the non-Fermi liquid regime is always covered by the superconducting phase, $\Delta \gg \mu_{NFL}$.

Comparing the previous scales shows that $\mu_{LD}$ has a loop suppression proportional to $g^2/(4\pi^2 v)$, while the non-Fermi liquid scale and the gap are both exponentially suppressed. In the weak coupling expansion, we will then always have $\mu_{LD} \gg \mu_{NFL}$ and $\mu_{LD} \gg \Delta$.\footnote{We are assuming that the UV parameters stay bounded as $\epsilon \to 0$, so that the theory can be connected to the gaussian fixed point.} The proposal is to accomplish this by
using (\ref{eq:Ddamped}) to compute interactions with the Fermi surface. This amounts to summing a special class of diagrams at every loop order, and it is necessary to understand under what conditions other effects at the same order can be neglected.
We now argue that this is actually consistent near $D=4$ spacetime dimensions in the weakly coupled limit $g^2 \ll 1$ (or $g^2/v \ll1$ in the case $v \gg 1$), after including additional logarithmic divergences from the running couplings.

The main question to address is whether higher loop effects can introduce corrections to the boson propagator that dominate over the one loop $\Pi(p)$ that we have taken into account. Analyzing these corrections obtains two types of effects. First, there are logarithmic corrections that come from the fermion self-energy and vertex subdivergences. In the RG approach these effects are rendered small by calculating the vacuum polarization in terms of the running gauge coupling and velocity. On the other hand, there are also higher order finite corrections which, similarly to the one loop $\Pi(p)$, are nonanalytic in $x$. Focusing for concreteness on the overdamped region $x \ll 1$, we find, by evaluating higher loop diagrams, that such effects have a well-defined Taylor expansion for small $x$, so they amount to a small correction of the one loop term, $M_D^2 |x| \to M_D^2 (1 + \mathcal O(g^2)+ \ldots) |x|$. Therefore, having resummed the leading one loop corrections responsible for the change from $z=1$ to $z=3$, higher loops appear to introduce only small deviations from this behavior.\footnote{It would be interesting to find an all-loop orders proof for the behavior of finite effects at small $x$. Also, establishing the control of the approximation at large $x$ is straightforward since this is the UV regime, where the theory is controlled by the tree level action.}

This should be contrasted with the situation in 2+1 dimensions where there is no weak coupling expansion in terms of $g$, and it is not clear which resummation would capture the dynamics of the theory~\cite{lee1, Metlitski:2010pd}. It would be interesting if an $\epsilon$--expansion of our results could help clarify this important problem, perhaps also in combination with a large number $N_f$ of fermion flavors, as in~\cite{Mross:2010rd}.

\subsection{Analytic and numeric approach}\label{subsec:results}

We will now re-analyze the fermion self-energy, but this time including the effects from damping via the one-loop resummed propagator (\ref{eq:Ddamped}). This section presents the basic procedure, while \S\S \ref{subsec:slow} and \ref{subsec:fast} summarize the numerical results for slow and fast fermions (which need to be studied separately).
Instead of dimensional regularization which we relied on before, it is easier numerically to work in $D=4$ with a hard cut-off $\Lambda$; the logarithmic divergence in $\Lambda$ is equivalent to an $\epsilon$ pole. The input couplings $g$ an $v$ are defined at the scale $\Lambda$, and the RG is obtained from $\Sigma(p)$ by varying the external fermion frequency or momentum. 

The starting point is the self-energy integral in terms of the Landau damped boson,
\be\label{eq:Sigma-damped1}
\Sigma(k_0, \vec k) =-\frac{g^2}{(2\pi)^4}\,\int \,\frac{dp_0\, d p_\perp \,d^2 p_\parallel}{p_0^2+p_\perp^2+p_\parallel^2+ \Pi(p_\mu)}\,\frac{1}{i(p_0+k_0)- v (p_\perp + k_\perp)}\,.
\ee
Given that $\Pi(p_\mu)$ depends on $|\vec p\,|$, instead of working with the components $p_\perp$ and $\pp$ it is more convenient to introduce radial coordinates, $p_\perp \equiv p \,\cos \theta$, $\pp \equiv p\,\sin \theta$, where here $p \equiv |\vec p\,|$. The one loop integral is then
\be\label{eq:Sigma-damped2}
\Sigma(k_0, \vec k) =-\frac{g^2}{(2\pi)^3}\int\, dp_0 \,dp\int_{-1}^1 d(\cos{\theta})\frac{p^2}{p_0^2+p^2+\Pi(\frac{p_0}{vp})}\,\frac{1}{i(p_0+k_0)- v (p\,\cos \theta + k_\perp)}\,.
\ee
In order to extract the logarithmic running of $\Sigma$ and compute the fermion anomalous dimension and velocity beta function, we will now compute in turn $\partial_{k_0} \Sigma$ and $\partial_{k_\perp} \Sigma$. The RG evolution will be obtained by varying the external frequency and setting $k_\perp=0$; the procedure for varying $k_\perp$ instead of $k_0$ is similar.

Integrating over $\cos \theta$ and expanding for small $k_\perp$ but keeping the whole $k_0$ dependence gives
\be\label{eq:sigma-damped3}
\Sigma(k_0, \vec k) \approx \frac{g^2}{4\pi^3 v}\,\int\frac{p \,dp\,dp_0}{p_0^2+p^2+\Pi(\frac{p_0}{vp})}\,\left(i \tan^{-1} \frac{vp}{p_0+k_0}+ \frac{v^2 p}{(p_0+k_0)^2 + v^2 p^2}\,k_\perp \right)
\ee
Let us discuss $\partial_{k_0} \Sigma$ first.
Since $\tan^{-1}(1/x) \approx \frac{\pi}{2}{\rm sgn}(x)$ as $x \to 0$, the derivative $\partial_{k_0} \Sigma$ is a sum of a regular piece (from the region $p_0 + k_0 \neq 0$) plus a singular contribution proportional to $\delta(p_0+k_0)$:
\be\label{eq:dwSigma}
\partial_{k_0}\Sigma(k_0)=d\Sigma_r(k_0)+d\Sigma_s(k_0)
\ee
where we have defined
\bea\label{eq:Sigmars}
d\Sigma_r(k_0)&=&-i \,\frac{g^2}{4\pi^3}\int_{p_0+k_0 \neq 0}\,\frac{p^2\,dp \,dp_0}{p_0^2+p^2+\Pi(\frac{p_0}{vp})}\,\frac{1}{p^2v^2+(p_0+k_0)^2}\nonumber\\
d\Sigma_s(k_0)&=&i \,\frac{g^2}{(2\pi)^2v}\int \,\frac{p \,dp}{k_0^2+p^2+\Pi(-\frac{k_0}{vp})}\,.
\eea
The delta function factor from the derivative acting on the discontinuity of $\tan^{-1}(1/x)$ was used to perform the $dp_0$ integral in $d\Sigma_s$. Notice that the singular contribution comes from the low frequency region, something that would be missed if we were to integrate over frequency shells. In fact, this singular contribution will be shown to dominate the low energy limit.

The calculation of $\partial_{k_\perp}\Sigma$ determines the running of the velocity and follows similar steps. The only difference is that there is no singular contribution as a function of $k_\perp$, as seen in (\ref{eq:sigma-damped3}). The regular part of $\Sigma(k_0,\vec{k})$ depends only on the combination $ik_0-vk_\perp$; the result is therefore simply
\be\label{eq:dlSigma}
\partial_{k_\perp}\Sigma(k_0)=iv\, d\Sigma_r(k_0)
\ee
From (\ref{eq:dlSigma}) it is clear that $d\Sigma_s$ is also the piece responsible for generating the Fermi-velocity flow in the un-screened regime. The flow as a function of external momenta can be computed in an analogous way, and we will quote the results below.

The numerical procedure is now the following: we evaluate the one-loop contribution to $\partial_{k_0} \Sigma$ and $\partial_{k_\perp} \Sigma$ from (\ref{eq:dwSigma}) and (\ref{eq:dlSigma}), as a function of external frequency $k_0$ and for different values of the Fermi-velocity $v$, with a cutoff scale $\Lambda$ for both the frequency and momenta.\footnote{The leading logarithmic dependence is not modified if the frequency and momenta cutoffs are different, as long as their ratio is fixed.} It is also convenient to work in units of the Debye mass. The regimes $v \ll 1$ and $v \gtrsim 1$ lead to different RG evolutions and will be discussed separately next.

\subsection{Slow fermions}\label{subsec:slow}

\begin{figure}[h!]
\begin{center}
\includegraphics[width=0.45\textwidth]{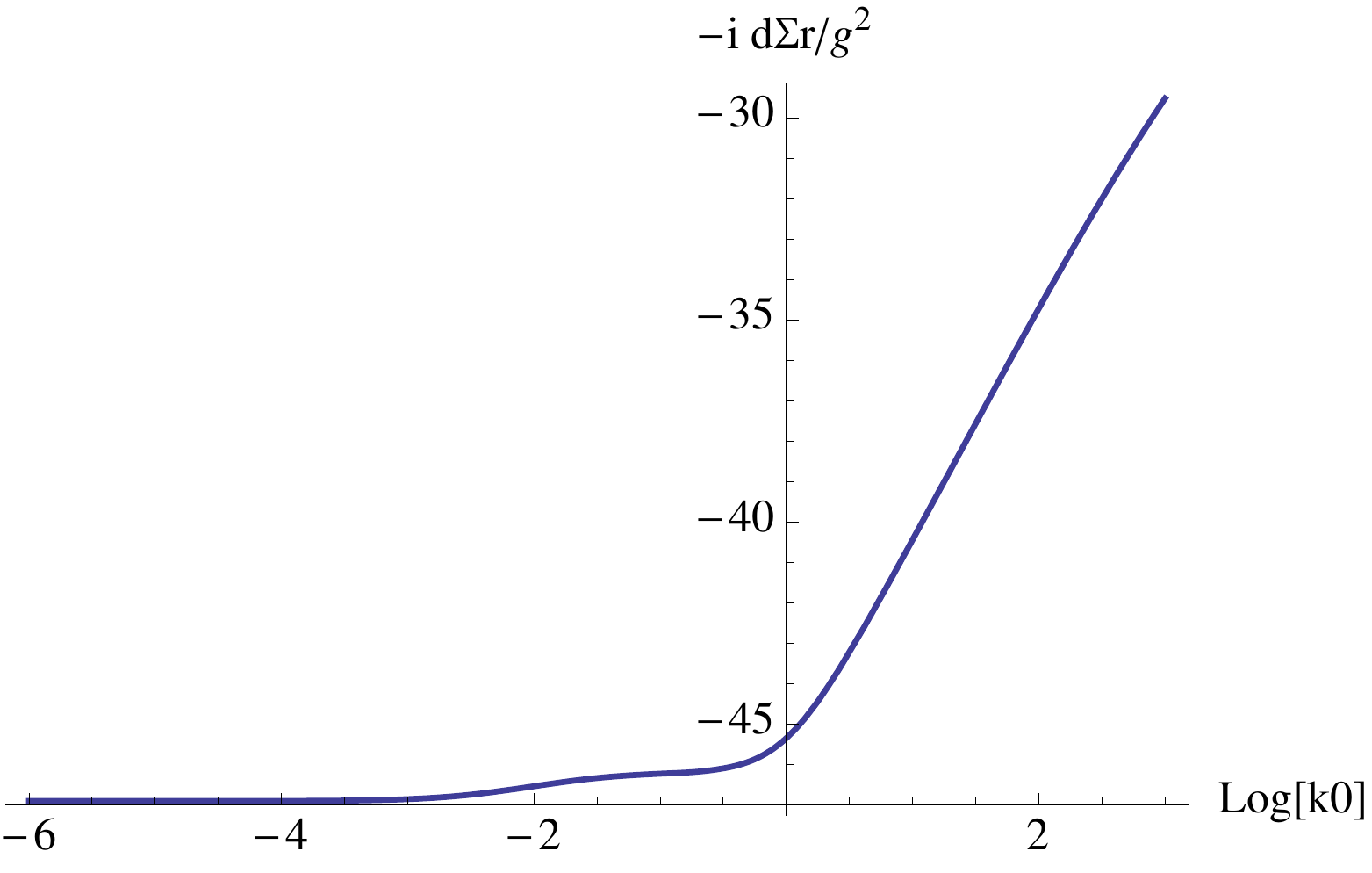}
\includegraphics[width=0.45\textwidth]{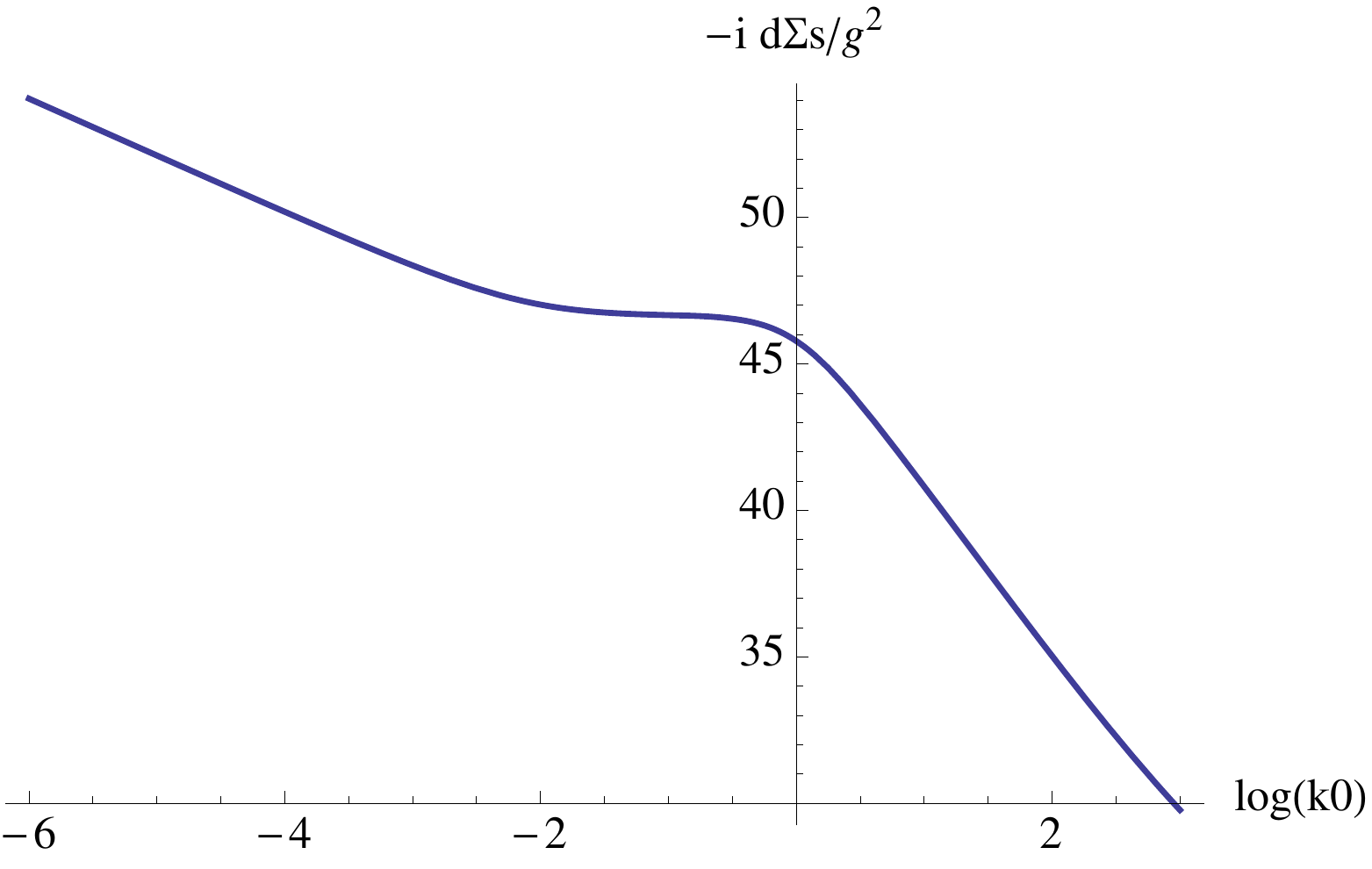}
\end{center}
\caption{\small{Plot of the regular and singular contributions to $-i \partial_{k_0} \Sigma$ as a function of $\log k_0$ for $v/c=0.01$ and $\Lambda=10^8$ in units of $M_D=1$.}}\label{fig:Sigmaslow}
\end{figure}

The numerical evaluation of $\partial_{k_0} \Sigma$ is shown in Fig.~\ref{fig:Sigmaslow}  with the choice $v/c=0.01$. 

First, in the regime of frequencies  $k_0\gg M_D $, $\Sigma$ depends linearly on $\log k_0$ and we have checked that this agrees with the undamped result (\ref{eq:dimreg_sigma}). This provides a  consistency check on our approach.

Next, the numeric evaluation shows that $\Sigma$ starts to deviate from the perturbative one loop answer at a scale $k_0 \sim M_D$
when Landau damping becomes important. This agrees with (\ref{eq:muslow-fast}) for slow fermions.
What follows below this scale is a very interesting crossover regime, where the effects of screening and damping transition from the dynamic $x \gg 1$ to the static $x \ll 1$ limits. This intermediate behavior extends roughly between the scales
\be
v M_D \ll k_0 \ll M_D\,,
\ee
and the self-energy has a rather nonlinear dependence on the frequency. Note that for very slow fermions this window of energies is parametrically large. The exit from this regime around the scale $k_0 \sim v M_D$ can be understood by examining $d\Sigma_s$ in (\ref{eq:Sigmars}). The theory enters the cross-over regime at $\mu_{LD}\sim M_D$, scale at which we can expand $\Pi(k_0,p)\sim M_D(1-\frac{|p|^2v^2}{3k_0^2}+...)$, and the integrand is peaked at $p\sim M_D$. As we lower $k_0$, $\Pi$ deviates from the dynamical limit, and there is an increased interplay between the dynamic and static effects. The break-down of the dynamical expansion is marked by $\frac{|p|^2v^2}{k_0^2}\sim 1$; taking $p\sim M_D$, we obtain the scale $k_0\sim M_D v$. Below this scale, the Landau damping transits into the static limit $\Pi\sim M_D^2\frac{k_0}{|p|v}$, which will begin to drive the over-damped dynamics.

Finally we come to the IR limit $k_0 < v M_D$. The regular contribution $d\Sigma_r$ is
approximately independent of the frequency, and only $d \Sigma_s$ is responsible for the logarithmic running.
In this limit, the contribution from $d\Sigma_s$ can be evaluated analytically [see (\ref{eq:Sigmars})]
\be\label{eq:dSigmaoverdamped}
d\Sigma_s(k_0)=i \,\frac{g^2}{4\pi^2|v|}\int \,\frac{p \,dp}{p^2+\frac{\pi}{2}M_D^2 \frac{|k_0|}{v p}}\approx  i\,\frac{g^2}{12\pi^2|v|}\log{\frac{\Lambda^3}{\frac{\pi}{2}M_D^2 k_0}}
\ee
which we checked agrees with the numeric answer. The cutoff in this expression is of order $M_D$, the momentum scale at which the overdamped approximation used in (\ref{eq:dSigmaoverdamped}) breaks down.
Furthermore, by (\ref{eq:dlSigma}) there is no logarithmic divergence in $\Sigma$ as a function of $k_\perp$. We conclude therefore that in this new IR scaling regime,
\be\label{eq:Sigma-resummed}
\Sigma(k_0, \vec k)\approx i\,\frac{g^2}{12\pi^2|v|}\,k_0\log\frac{\mu}{k_0}\,,
\ee
where $\mu$ is the RG scale at which the physical couplings $g$ and $v$ are evaluated.

There is also a one loop renormalization of the vertex. By direct calculation or from (\ref{eq:Sigma-resummed}) together with the identity (\ref{eq:ward_identity}), the result is
\be
\Gamma(k;q) =\frac{g^3}{12\pi^2|v|}\,\frac{iq_0}{iq_0-v q_\perp}\,\log\frac{\mu}{q_0}\,.
\ee
We will treat this nonlocal divergence as in (\ref{eq:Gamma-interpret}).

We should stress that even though we have used the nonlocal boson propagator to calculate the backreaction on the Fermi surface, the resulting fermion self-energy is well-behaved. Apart from the crossover regime where the system adjusts itself to static and dynamic damping effects, we have found a logarithmic running in the UV that reproduces the perturbative answer, and a new regime below the scale $v M_D$. The running here is still logarithmic, so this is a controlled correction to the classical theory. However, it has important differences with the one loop result (\ref{eq:dimreg_sigma}), and we will discuss its consequences on the long distance theory in \S \ref{sec:IR}.

\subsection{Fast fermions}\label{subsec:fast}

\begin{figure}[h!]
\begin{center}
\includegraphics[width=0.45\textwidth]{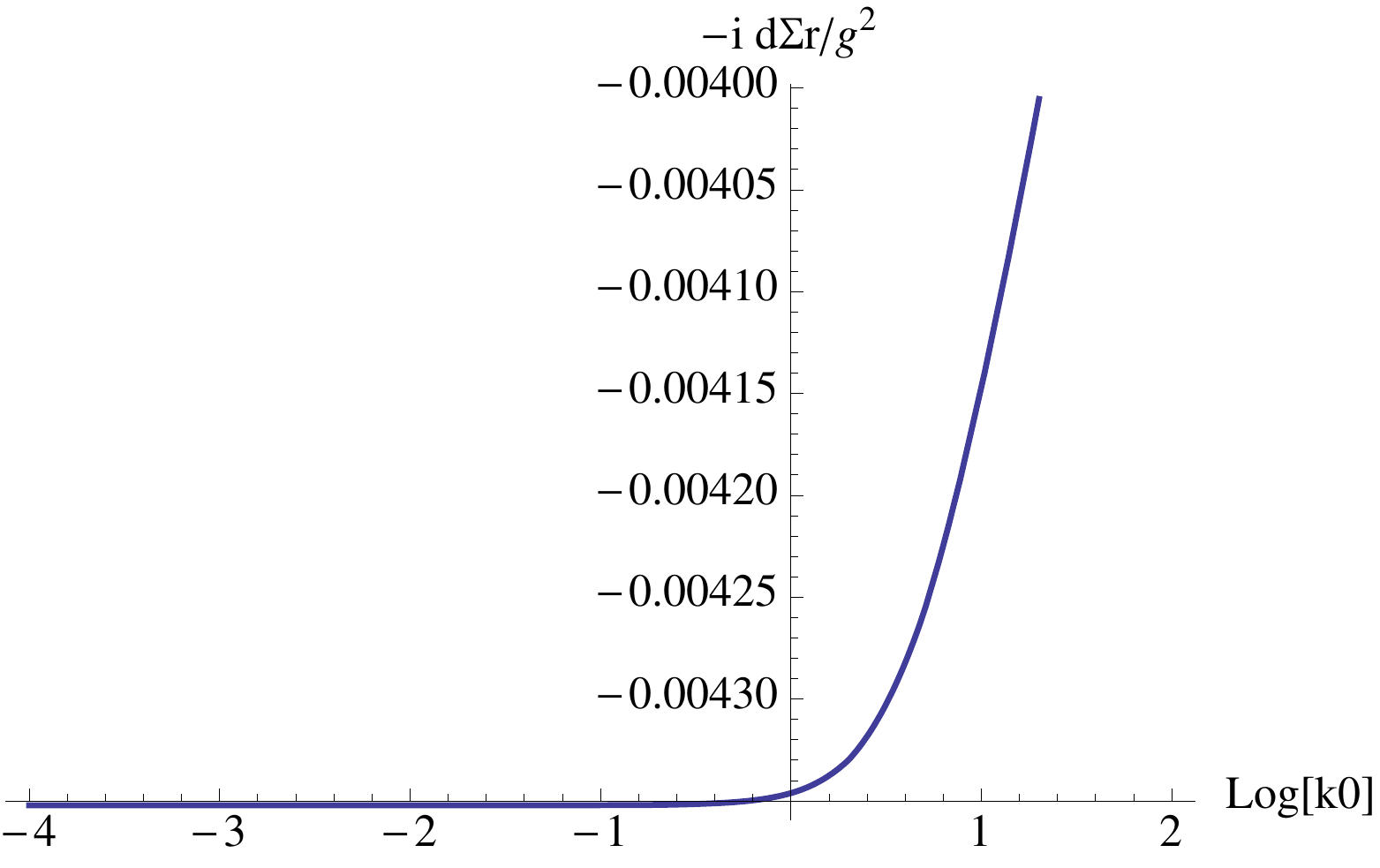}
\includegraphics[width=0.45\textwidth]{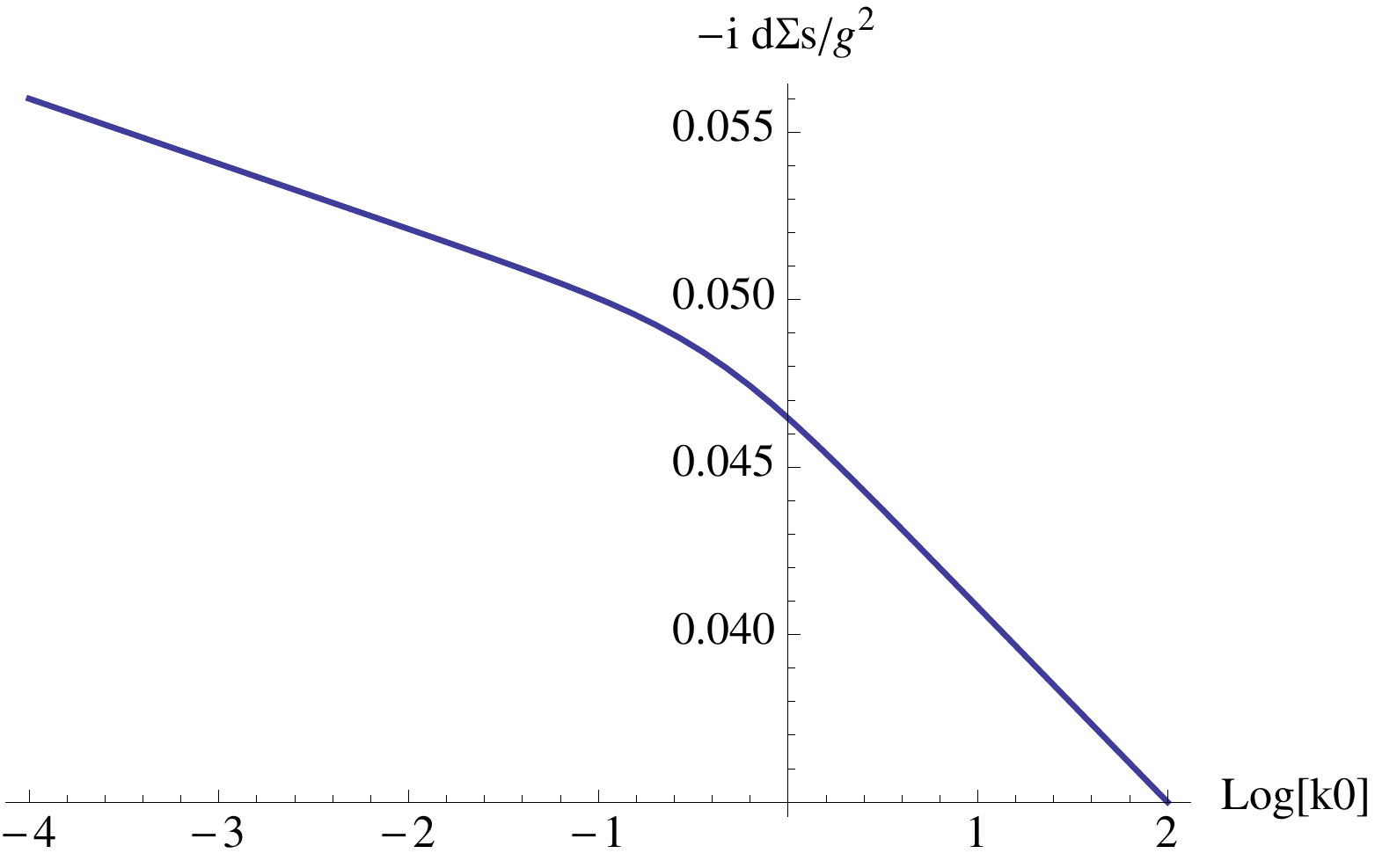}
\end{center}
\caption{\small{Plot of the regular and singular contributions to $-i \partial_{k_0} \Sigma$ as a function of $\log k_0$ for $v/c=10$ and $\Lambda=10^8$ in units of $M_D=1$.}}\label{fig:Sigmafast}
\end{figure}

Let us now repeat the above procedure for fast fermions $v \gg c$. As an example,
the numerical evaluation of $\partial_{k_0} \Sigma$ is shown in Fig.~\ref{fig:Sigmafast}  with the choice $v/c=10$. 

Comparing both cases $v \ll c$ and $v \gg c$, we see two main differences. First, for fast fermions the one loop approximation is valid up to a parametrically lower scale $k_0 \sim M_D/v^{1/2}$ (as opposed to $k_0 \sim M_D$ for slow fermions). Taking $v \gg c$ helps to power-law suppress Landau damping. The second difference is that now the crossover regime between the UV and overdamped limits occurs rather quickly, instead of the broad crossover discussed above. The reason for this can be traced back to the different behaviors (\ref{eq:Piscalarlimit}). As we have argued before, the backreaction of Landau damping on the Fermi surface starts to become important when $p_0^2+ \vec p^{\,2} \sim \Pi(p_\mu)$ for a nearly on-shell scalar $p_0 \sim |\vec p\,|$. For a fast fermion, this means that $x=p_0/(v|\vec p\,|) \ll 1$, for which $\Pi \sim M_D^2/v$. This reproduces the scale $k_0 \sim M_D/v^{1/2}$ observed numerically. 

Therefore, there is now no intermediate regime where the theory transitions between dynamic and static limits. Finally, the deep IR behavior is the same for both slow and fast fermions, given in (\ref{eq:Sigma-resummed}).

\section{Low energy dynamics}\label{sec:IR}

In this section we analyze the physical consequences of our previous results. After summarizing in \S \ref{subsec:summary} the three different regimes of the theory, in \S \ref{subsec:IR} we focus on the theory in the overdamped limit. We will find that it describes two sectors --the boson and the Fermi surface-- with different dynamical exponents, and we show that perturbation theory reorganizes in terms of an effective coupling $g^2/v$ which becomes strong in the IR due to the renormalization of the Fermi velocity.

\subsection{Summary of regimes}\label{subsec:summary}

It will be useful to first summarize the results of \S \ref{sec:LD} in the following three energy ranges.

\textit{UV undamped limit}: For energies above the Landau damping scale $\frac{\mu_{LD}}{M_D} \approx \left(v^{-1}\,\tan^{-1}(v) \right)^{1/2}$ the fermion self-energy calculated using the corrected boson propagator agrees with the one loop result and the dynamics of \S \ref{sec:oneloop} applies. The theory starts to flow to $v \to 0$, and the fermion acquires a nonzero anomalous dimension (\ref{eq:one-loop-beta}). As we discussed before, for $\epsilon >0$ the coupling is relevant; in the finite $N$ generalization of \S \ref{subsec:finiteN} it is also possible to find a fixed point for $g$.

\textit{Crossover regime}: For systems with $v \ll c$, Landau damping cuts off the perturbative flow at $\mu\sim M_D$. Our analysis then reveals an interpolating regime that can extend over a parametrically large window of scales,
\be\label{eq:intermediate-window}
v\,M_D \ll \mu \ll M_D\,.
\ee
The physics here is controlled by the ratio $x=p_0/(v |\vec p\,|)$,
and this region describes the interpolation between two different UV and IR scaling regimes. 

In models with $v \gtrsim c$, Landau damping is suppressed and cuts off the perturbative flow at the lower scale of $\mu\sim M_D v^{-1/2}$. The interpolating range for the case of $v\ll c$ collapses to a rapid crossover between the undamped and overdamped theories. It would be interesting to study this regime in more detail, and detect potentially observable consequences.

\textit{Overdamped limit}: Finally, we come to the dynamics below the scale $v M_D$ for slow fermions, or $M_D/v^{1/2}$ for fast fermions. The low energy theory exhibits two important features. First, the boson propagator is now controlled by Landau damping, with a $z=3$ dynamical exponent. The RG evolution that we have constructed then explains how $z=3$ emerges in the IR, and connects it with the perturbative UV theory with $z=1$, via the nontrivial crossover described above. 

The other aspect is that the IR theory organizes as a perturbative expansion in the effective coupling
\be\label{eq:alphadef}
\alpha \equiv \frac{g^2}{12\pi^2 |v|}\,,
\ee
both for slow and fast fermions. This can be seen by redefining the momenta to set $v=1$ in the fermion dispersion relation. Further rescaling the fields makes the fermion canonical, changes the boson kinetic term to $L \supset \phi (v^2 p_0^2 + \vec p^{\,2} )\phi$, and the Yukawa coupling to $g/v^{1/2}$. The tree level frequency dependence in the boson kinetic term is subleading in the overdamped regime, so in this limit the theory depends only on the coupling $g/v^{1/2}$, and this explains (\ref{eq:alphadef}). In what follows we will study in more detail the dynamics in this range.

\subsection{IR dynamics}\label{subsec:IR}

We now take a closer look at the IR theory in the over-damped regime. It contains two dynamical exponents: the boson has $z_b=3$, while for the fermion $z_f=1$. The two exponents compete inside loop integrals, and one has to determine which one dominates. One approach is to integrate out the Fermi surface to obtain an effective action for the boson, and then construct a scaling theory around the $z=3$ exponent~\cite{hertz}. However, there are various reasons why this is not the whole story. First, this action is nonanalytic, due to the light quasiparticles from the Fermi surface. One manifestation is in the logarithmic corrections to Landau damping that we discussed in \S \ref{subsec:Piresum}. Another related effect is that the $\phi^4$ interaction calculated in \S \ref{subsec:boxboson} also has a singular momentum dependence at the fermion dispersion relation. Furthermore, the Hertz theory is expected to break down at the scale of superconductivity.

For these reasons, it is necessary to focus as well on the scaling determined by the Fermi surface, which is also the appropriate scaling for superconductivity. Let us now consider the beta functions of the theory assuming the perturbative $z=1$ scaling. Combining (\ref{eq:Sigma-resummed}) with the tree level kinetic term gives an anomalous dimension $\gamma_\psi=g^2/(24\pi^2|v|)$, and $\delta_v=0$ in the notation of (\ref{eq:scalarbetafc}). The velocity flows to zero with a rate proportional to $\gamma_\psi$. As we argued before, the effective coupling in the low energy theory is actually $\alpha$ in (\ref{eq:alphadef}) and not just $g$, and the beta functions become
\be\label{eq:NFL-overdamped1}
\gamma_\psi= \frac{\alpha}{2}\;,\;\beta_v=\alpha\,v\;,\;\beta_\alpha=-\epsilon \alpha - \alpha^2\,.
\ee
We see that even if $\epsilon=0$, the interaction is driven to strong coupling in the IR, which is a consequence of the running velocity $v \to 0$ in combination with Landau damping. This is an interesting effect: while in relativistic theories in 3+1 dimensions nonabelian gauge interactions are needed to have asymptotic freedom, in the nonrelativistic case quantum effects can give $\beta_\alpha <0$ already in the Yukawa theory we are considering. For the case of a gauge field instead of a scalar, the analog effect makes the coupling IR free, and it is possible to find a weakly coupled fixed point by balancing the tree level and one loop contributions~\cite{nortonNFL}. This fixed point does not exist in the scalar theory because the one loop contribution is $\beta_\alpha <0$.

The strong coupling limit is not reached, however. The reason is that the superconducting (SC) instability, which sets in at a scale $\Delta \sim e^{-1/\sqrt{\alpha}} \Lambda$, is much larger than the non-Fermi liquid scale $\mu_{NFL} \sim e^{-1/\alpha} \Lambda$ in our perturbative regime. Therefore the non-Fermi liquid phase is covered by a SC dome. One interesting question is whether it is possible to have the quantum critical point not completely covered by the superconductor phase, as displayed in various strongly correlated systems.\footnote{A recent analysis of this in a different class of models appeared in~\cite{Metlitski:2014zsa}.} We will find that this is indeed possible in a nonabelian generalization of the theory, to which we turn next.

\section{Phase structure at finite $N$}\label{sec:phases}

Finally, we will consider a natural extension of the theory we have analyzed so far, to allow for nonabelian global symmetries $SU(N_c) \times SU(N_f)$. The first generalizes spin rotations, while $N_f$ can arise in multichannel systems. Technically, our previous results extend readily to this case, but we will find that the low energy physics can be quite different. Our motivation for this is that varying $N_c$ and $N_f$ will allow us to access different phases of the theory, some of which could have applications to strongly correlated electronic systems. After briefly discussing in \S\ref{subsec:finiteN} the extension to finite $N$, in \S \ref{subsec:phases} we present a preliminary analysis of the IR phases of the theory. A rich phenomenology emerges for different values of $N_c$, with the possibility of having the non-Fermi liquid behavior drive superconductivity, something that could be of relevance for high $T_c$ materials.

\subsection{Finite $N$ generalization}\label{subsec:finiteN}

The theory for general $(N_c, N_f)$ has the field content
\be
\begin{tabular}{c|cc}
&$SU(N_c)$&$SU(N_f)$\\
\hline
&&\\[-12pt]
$\phi_{ij}$ & adj+1 & $1$  \\
$\psi_{ia}$ & $\Box$ & $\overline \Box$
\end{tabular}
\ee
The invariant Yukawa interaction is $L \supset g \phi_{ij}\psi^\dag_{ia} \psi_{ja}$. Let us for simplicity consider large $N_f$ and $N_c$, with $N_f/N_c$ fixed. In order to have a well-defined large $N$ limit, we should also keep
\be\label{eq:'tHooft}
\t \alpha \equiv \frac{g^2 N_c}{12\pi^2|v|}
\ee
fixed, the analog of the 't Hooft coupling in our model. 

The nonabelian symmetries introduce two main modifications in the one loop results. First, Landau damping is rescaled by a factor of $N_f/N_c$,
\be
\Pi(x) = 6\,\frac{N_f}{N_c}\,\t \alpha\, k_F^2\,x \tan^{-1}(1/x)\,,
\ee
and hence $N_f/N_c$ changes the scale at which $\Pi(x)$ starts to dominate. In particular, if $N_c \gg N_f$ Landau damping is suppressed, while in the opposite limit $N_f \gg N_c$ it is enhanced. In the latter case, the 't Hooft coupling should be defined in terms of $N_f$ instead of $N_c$ so as to have a well-defined large $N$ limit. As long as $N_f/N_c$ is finite and no parameters become exponentially large $\sim e^{1/\t \alpha}$ for $\t \alpha \to 0$ (the perturbative limit around which we are expanding), the Landau damping scale is larger than the one loop $\mu_{NFL}$, and the resummed boson propagator has to be used. The justification of this is similar to that given in \S \ref{subsec:Piresum}. 

The second effect is that the anomalous dimension and vertex contributions to $\beta_g$ no longer cancel. The first is multiplied by the $SU(N_c)$ Casimir $T^A T^A=C_2(\Box)$, while for the second the color factor is $T^B T^A T^B = \left(C_2(\Box)-\frac{1}{2} C_2(\text{adj}) \right)T^A$. At large $N_c$ the vertex contribution is then suppressed. The resulting beta functions in the overdamped regime $\mu< \mu_{LD}$ are
\be\label{eq:NFL-overdamped2}
\gamma_\psi= \frac{\t \alpha}{2}\;,\;\beta_v= \t \alpha \,v\;,\;\beta_{\t \alpha}= - \epsilon\,\t \alpha+ \t \alpha^2\,,
\ee
which extend (\ref{eq:NFL-overdamped1}) to $N_c>1$. We see that for $N_c>1$ the one loop contribution to $\beta_{\t \alpha}$ changes sign; now quantum effects tend to screen the coupling in the IR. As a result, and unlike the $N_c=1$ case, 
the nonabelian theory admits a non-Fermi liquid fixed point at
\be\label{eq:NFLoverdamped}
\t \alpha_* = \epsilon\;,\;\gamma_\psi=\epsilon/2\,,
\ee
where both the velocity and the quasiparticle residue flow to zero with a power-law form
\be\label{eq:vZ}
v(\omega) = v(\mu)\, \left(\frac{\omega}{\mu} \right)^{\epsilon}\;,\;Z(\omega) = Z(\mu)\,\left(\frac{\omega}{\mu} \right)^{\epsilon}\,.
\ee
This non-Fermi liquid (NFL) fixed point is under perturbative control, and uses Landau damping in an essential way as discussed in \S \ref{subsec:IR}. We should also recall that the full theory is not critical, since the bosonic sector follows a different $z=3$ scaling.

\subsection{Preliminary analysis of IR phases}\label{subsec:phases}

We will end our analysis with a preliminary discussion of the IR phases. We have found that for $N_c>1$ the system has a NFL phase coupled to a $z=3$ boson. Now we need to estimate the scale of the superconducting instability. Since $\langle \psi_{ia}(x) \psi_{ja}(x)\rangle$ is not a singlet of $SU(N_c)$, the one loop contribution is non-planar; the gap is hence not enhanced by $N_c$, and the dependence is still given by (\ref{eq:gapscale}):
\be
\Delta \sim e^{-\sqrt{N_c}/\sqrt{\t\alpha_*}} \Lambda\,.
\ee
The dominant $l=0$ angular momentum mode for the BCS condensate forms in the antisymmetric of $SU(N_c)$.
This should be compared with the NFL scale $\mu_{NFL} \sim e^{-1/\t \alpha_*}\Lambda$ of (\ref{eq:NFLoverdamped}). For small $N_c$, the NFL region is completely covered by the superconducting dome (SC), but as $N_c$ increases to $N_c \gtrsim \alpha_*^{-1}$ the NFL regime sets in before the superconducting instability.

\begin{figure}[h!]
\begin{center}
\includegraphics[width=0.80\textwidth]{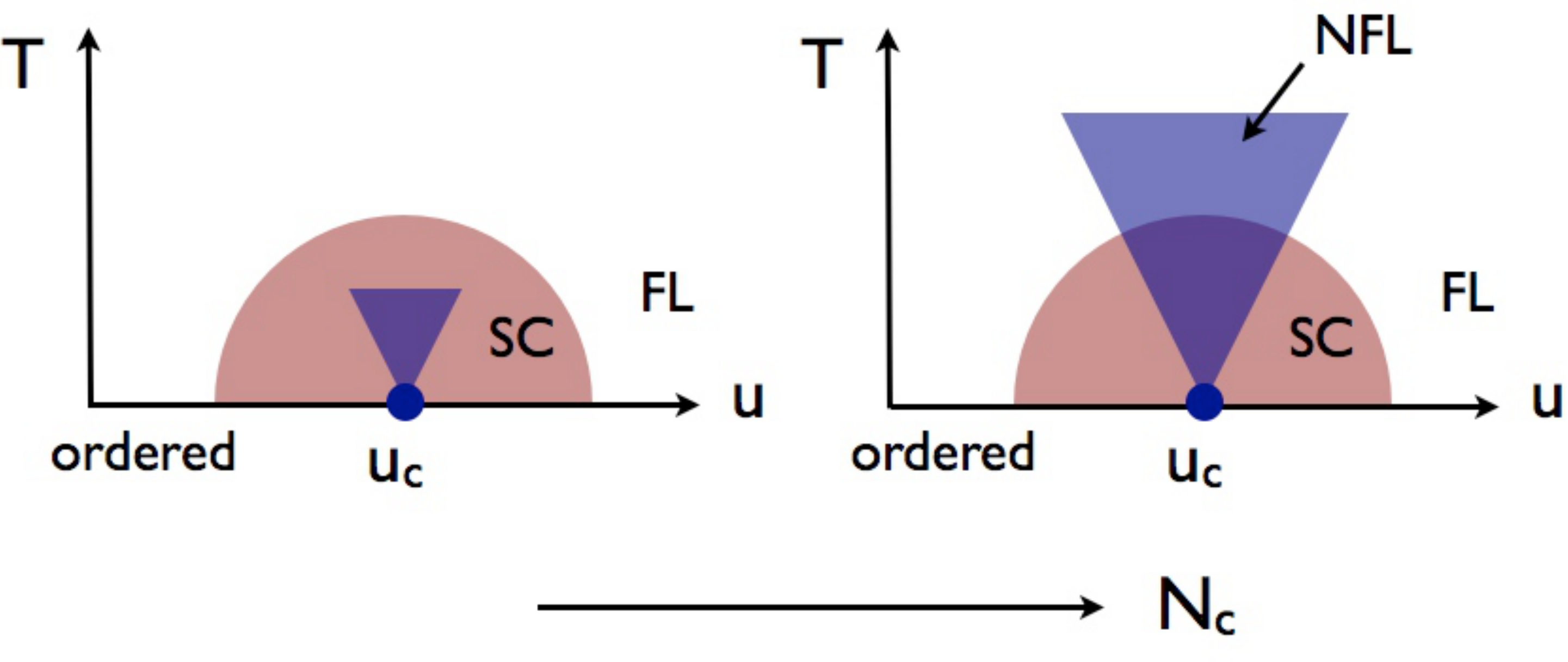}
\end{center}
\caption{\small{Schematic representation of the phase diagram as a function of the parameter $u$ that tunes to criticality and temperature, and different values of $N_c$. For small $N_c$ the superconducting region (SC) completely covers the non-Fermi liquid (NFL) region. As $N_c$ increases, the NFL region extends beyond the SC phase}}\label{fig:phases}
\end{figure}

Ignoring for the moment other possible instabilities, the phase diagram of the theory is illustrated schematically in Figure \ref{fig:phases}. Here $u$ is a control parameter that tunes the boson to criticality, in the way described in \S \ref{subsec:vacuumpol}.

It is very encouraging that already the simple theory considered in this work, in its weak coupling expansion, displays such a rich phenomenology. This model can be
potentially relevant for various strongly correlated systems, both in $d=3$ space dimensions, or by developing higher orders of the $\epsilon$--expansion to try to model quasiplanar systems. Changing the parameter $N_c$, the generalization of $SU(2)$ spin, leads to a very different interplay between superconductivity and quantum criticality. It is intriguing that compounds dominated by single-band or multi-band interactions also display different phase diagrams~\cite{shibauchi}.

The prediction of a NFL regime above the SC dome for moderately large $N_c$ should be highlighted.\footnote{Estimating the critical value of $N_c$ for which the NFL extends above the SC dome requires taking into account order one factors which we haven't done here, especially in the calculation of the gap. For $\epsilon \sim 0.1$, already for $N_c \sim 5$ NFL effects start to become important near $\Delta$.} Over a broad range of initial parameters the theory flows to this regime, which incorporates Landau damping backreaction on the Fermi surface, and where the velocity and quasiparticle residue flow to zero as described in (\ref{eq:vZ}). This robust prediction makes this model an ideal candidate to study how quantum criticality can drive superconductivity and strange metallic behavior, a subject of importance for understanding high $T_c$ materials. 

Finally, it will clearly be important to study other possible IR phases of the theory. We have focused on the SC instability, but there can be other instabilities such as stripe order, either from the Fermi liquid or from the boson. For instance, at large enough $N_c/N_f$ a charge density wave (CDW) instability in the Fermi surface can arise (as in QCD~\cite{CDW}); this can have strong effects since the CDW may not be suppressed by large $N_c$. It would also be interesting to analyze the fate of the boson, and the interplay between its ordered phase and the NFL, SC and CDW phases. We hope to return to these points in the future.

\section{Conclusions and future directions}\label{sec:concl}

Motivated by evidence for quantum phase transitions and NFL behavior, as well as by theoretical questions on the renormalization of nonrelativistic QFTs, in this work we have studied the coupled theory of a critical boson and a Fermi surface in $D=4-\epsilon$ spacetime dimensions. Using the $\epsilon$--expansion and in a perturbative limit, we included the full backreaction of Landau damping, both on the scalar and the fermions, and mapped the RG evolution across the different relevant energy scales.

We found a new crossover regime, connecting the one loop and overdamped behaviors, where static and dynamic damping effects are important, and there is no scaling symmetry. Below the crossover, the theory is described by a $z_b\approx3$ bosonic sector (which includes nonanalytic corrections from the Fermi surface), interacting with a $z_f\approx1$ NFL. After generalizing the model to allow for a nonabelian global symmetry (the extension of spin rotations), we identified a NFL fixed point over a large parameter space of the theory. The critical behavior is characterized by a Fermi velocity and quasiparticle residue $Z$ that flow to zero as $\omega^\epsilon$. An important property of the theory is that by increasing the number of bosonic flavors, it is possible to interpolate between a phase where the NFL is covered by the SC instability, and the case where the NFL becomes important firsts and affects the SC.

Let us end by discussing some of the future directions of research. At a more conceptual level, the non-local vertex renormalization found in this paper requires a detailed understanding, something which we hope to address in the future~\cite{liamfuture}. The non-locality is associated with the Fermi velocity running and the presence of low energy modes across the Fermi surface, so this may be a more generic phenomenon. This raises conceptual challenges in making sense of, and eventually constructing, low-energy effective theories for nonrelativistic QFTs. 
Secondly, it would be interesting to extend the analysis to higher orders in $\epsilon$, with the hope of capturing some of the dynamics of quasiplanar systems in a controlled framework. Finally, it will also be important to fully explore the IR phases of such system, and investigate how various orders compete in different regions of the phase diagram. With bosonic fluctuations arising both from the scalar field and as collective modes of the Fermi surface, it may be possible to realize phenomena such as stripe order and Fermi surface reconstruction. It would also be interesting to develop more realistic models, and check if the spin number $N_c$ that played an important role in the IR phases of our theory can also be related to multiband effects.

\section*{Acknowledgments}

We would first like to thank  L. Fitzpatrick, S. Hartnoll, S. Kachru, M. Mulligan and S. Raghu for extensive discussions on non-Fermi liquids and on results related to this work. We also thank
L. Fitzpatrick,
S. Hartnoll,
S. Kachru,
M. Mulligan,
S. Raghu, and
S. Sachdev
for useful comments on the manuscript.
G.T.~is supported in part by the National Science Foundation under grant no.~PHY-0756174. H.W.~ is supported by a Stanford
Graduate Fellowship.

\appendix

\section{Conventions and useful formulas}\label{app:useful}

\subsection{Field theory conventions}

We work in euclidean signature. Our convention for Fourier transforms is $f(x) \sim \int_p e^{-i px} f(p)$, and the path integral measure is $e^{-S}$, with the action defined in the main text. Let us focus first on the fermions, which require a bit more of care than bosons. Under Fourier transform, the fermion kinetic term close to the Fermi surface becomes $L_{kin}= -i \psi^\dag(p) (i p_0 - vp_\perp) \psi(p)$, where $\vec p = \hat n (k_F+p_\perp)$. Recalling that for a fermionic path integral
$
Z= \int D\psi D \psi^\dag e^{\psi^\dag_\alpha K_{\alpha \beta} \psi_\beta}
$
the propagator is $\langle \psi_\alpha \psi^\dag_\beta \rangle = - K^{-1}_{\alpha \beta}$, in our case we have the Green's function
\be
\langle \psi(p) \psi^\dag(p) \rangle= G_F(p_0, \vec p)= - \frac{1}{i p_0- v p_\perp}\,.
\ee
For a scalar field,
\be
\langle \phi(p) \phi(-p) \rangle =  D(p)= \frac{1}{p_0^2 + \vec p^{\,2}}\,.
\ee
When calculating quantum corrections, we will add a superindex `$(0)$' to these tree-level propagators.

The one loop fermion self-energy $\Sigma$ generated by the interaction with a a scalar or a gauge field is given by
\be\label{eq:Sigmadef}
G_F^{(1)} = G_F^{(0)}+G_F^{(0)}\,\Sigma\,G_F^{(0)} + \ldots = \frac{1}{[G_F^{(0)}]^{-1}-\Sigma}\,,
\ee
where
\be\label{eq:Sigmaoneloop}
\Sigma(p) =\mu^\epsilon g^2\, \int \frac{d^Dq}{(2\pi)^D}\,G_F^{(0)}(p+q) \,D^{(0)}(q)\,.
\ee

There is a similar one loop correction for the boson induced by the fermion loop:
\be\label{eq:Pidef}
D^{(1)}= D^{(0)}-D^{(0)} \Pi D^{(0)} + \ldots = \frac{1}{[D^{(0)}]^{-1}+\Pi}
\ee
with
\be\label{eq:Pioneloop}
\Pi(p)=\mu^\epsilon g^2\, \int \frac{d^Dq}{(2\pi)^D}\,G_F^{(0)}(q) \,G_F^{(0)}(p+q)\,.
\ee
The extra minus sign here comes from the fermion loop. The loop integral can be decomposed into an integral over the momentum normal to the Fermi surface times the remaining $(d-1)$--dimensional angular part.

Lastly, consider the correction to the interaction $L\supset \mu^{\epsilon/2}\,g  \psi^\dag(q+k) \phi(q) \psi(k)$. Writing the quantum vertex as (there is an overall minus sign from $e^{-S}$)
\be
-\langle\psi^\dag(k+q) \phi(q)  \psi(k) \rangle_\text{amp}= \mu^{\epsilon/2}\left(g+\Gamma(k;\,q) \right)
\ee
the one loop contribution is given by
\be\label{eq:vertexoneloop}
\Gamma(k;\,q)=\mu^\epsilon g^3 \int \frac{d^Dp}{(2\pi)^D} D^{(0)}(p-k)\,G_F^{(0)}(p)\, G_F^{(0)}(p+q)\,.
\ee

\subsection{Some useful integrals}\label{subsec:integrals}

Our main starting point will be the integral
\be
I_n \equiv \int dp_0 \, dp_\perp \,d^{d-1} p_\parallel \,\frac{1}{(A\, p_0^2 + B\,p_\perp^2 + C \, p_\parallel^2+\Delta)^n} 
= \frac{\pi^{\frac{d+1}{2}}\Gamma\left(\frac{2n-d-1}{2} \right)}{\Gamma(n)}\,\frac{1}{\sqrt{ABC^{d-1}\Delta^{2n-d-1}}}
\ee
where the coefficients in the denominator are positive and we assume $d<2n+1$ so that the integral converges. Taking derivatives of this expression with respect to $A$, $B$, $C$ or $\Delta$ obtains other integrals that are also useful in our computations. For instance, a derivative with respect to $A$ gives
\be
\int d p_0 \, dp_\perp \,d^{d-1} p_\parallel \,\frac{p_0^2}{(A\, p_0^2 + B\,p_\perp^2 + C \, p_\parallel^2+\Delta)^n}=\frac{\pi^{\frac{d+1}{2}}\Gamma\left(\frac{2n-d-3}{2} \right)}{2\Gamma(n)}\,\frac{1}{\sqrt{A^3BC^{d-1}\Delta^{2n-d-3}}}\,.
\ee
We will be interested in the case $d=3-\epsilon$ with $\epsilon \ll 1$, both for dimensional regularization and the $\epsilon$--expansion.

For instance, using this formula (and the usual Feynman parameters) we can derive the following integrals that enter the calculation of the self-energy, for small $\epsilon$:
\be\label{eq:I1}
\int \frac{d^Dp}{(2\pi)^{D}}\,\frac{1}{i(p_0+k_0)- v (p_\perp+k_\perp)}\,\frac{1}{p_0^2+p_\perp^2+p^2}= -\frac{1}{\epsilon} \frac{1}{4\pi^2}\,\frac{i k_0+ {\rm sgn}(v) k_\perp}{1+|v|}+\mc O(\epsilon^0)\,,
\ee

\subsection{Vertex correction in dimensional regularization}\label{subsec:appVertex}

This Appendix presents the calculation of the vertex correction with external boson momenta $(q_0, q_\perp)$ in dimensional regularization, which is proportional to the integral
\be
I=\int \frac{d^{D}p}{(2\pi)^D}\,\frac{1}{p_0^2+p_\perp^2+\pp^2}\,\frac{1}{i(q_0+p_0)-v(q_\perp+p_\perp)}\,\frac{1}{i p_0-v p_\perp}\,.
\ee
The dependence on the external fermion momentum is continuous, so here we have set $k_0=0$, $\vec k = \hat n k_F$, and we have decomposed the internal momenta in components perpendicular and parallel to $\hat n$.

First, the denominators are combined using Feynman parameters, and the loop momenta are shifted in order to complete squares in the denominator:
\be
I=2\int_0^1dx\int_0^{1-x}dy\int \frac{d^D p}{(2\pi)^D}\frac{(i(q_0+ p_0-\delta_0)+v(q_\perp+p_\perp-\delta_\perp))(i(p_0-\delta_0)+v(p_\perp-\delta_\perp))}{\left[p_0^2+\left(1-(1-v^2)(x+y)\right)p_\perp^2+(1-x-y)\pp^2+\Delta(x,y)\right]^3}\,.
\ee
The original loop momenta were shifted by $p_0 \to p_0+\delta_0,\,p_\perp \to p_\perp+\delta_\perp$, with $\delta_0=x q_0,\,\delta_\perp=\frac{x v^2 q_\perp}{1-(1-v^2)(x+y)}$, and $\Delta(x,y)$ is independent of momenta. 

Eliminating the odd terms in the numerator and performing the loop integral using \S \ref{subsec:integrals} obtains
\be
I=\frac{\pi^3\csc{\frac{\pi\epsilon}{2}}}{2(2\pi)^4}\int_0^1dx\int_0^{1-x}dy\frac{-(1-v^2)(1-x-y)^{\frac{\epsilon}{2}}}{(1-(1-v^2)(x+y))^{\frac{3}{2}}\Delta(x,y)^{\frac{\epsilon}{2}}}+\frac{A(x,y)(1-x-y)^{-1+\frac{\epsilon}{2}}\epsilon}{(1-(1-v^2)(x+y))^{\frac{1}{2}}\Delta(x,y)^{1+\frac{\epsilon}{2}}}
\ee
and we have defined $A(x,y)=\delta_0(q_0-\delta_0)-\delta_\perp v^2(q_\perp-\delta_\perp)$. The first term here is UV divergent, and now we find an additional contribution from the second term. Naively it is suppressed by $\epsilon$; however in the limit when $y \to 1-x$, the integral of the numerator is made convergent by $\epsilon$: the would-be logarithmic singularity is replaced by $1/\epsilon$, which cancels the $\epsilon$ suppression in front. The leading $\epsilon$ dependence is then
\bea
I=\frac{\pi^3\csc{\frac{\pi\epsilon}{2}}}{2(2\pi)^4}\left(-\frac{2(1-|v|)}{(1+|v|)|v|}+\int_0^1dx\frac{2A(x,1-x)}{\Delta(x,1-x)|v|}+\mc O(\epsilon)\right)\,.
\eea

Now one can check that: 
\be
A(x,1-x)=x(x-1)(iq_0+vq_\perp)^2\;,\;
\Delta(x,1-x)=x(1-x)(q_0^2+v^2 q_\perp^2)\;,
\ee
so $\frac{A(x,1-x)}{\Delta(x,1-x)}$ is independent of $x$,  and we have
\be\label{eq:vertex}
I(q_0, q_\perp)=\frac{1}{\epsilon}\frac{1}{4\pi^2(1+|v|)}\frac{iq_0+\sv q_\perp}{i q_0-v q_\perp}+\mc O(\epsilon^0)\,.
\ee 

\bibliographystyle{JHEP}
\renewcommand{\refname}{Bibliography}
\addcontentsline{toc}{section}{Bibliography}
\providecommand{\href}[2]{#2}\begingroup\raggedright

\end{document}